\documentclass[
preprint,
 amsmath,amssymb,
 aps,
]{revtex4-2}

\usepackage{graphicx}
\usepackage{dcolumn}
\usepackage{bm}
\usepackage[version=3]{mhchem}
\usepackage{float}
\usepackage{makecell} 
\usepackage{longtable}
\usepackage{booktabs}

\preprint{}

\begin{document}

\title{Intermolecular Interactions between Polyethylene, Water, and Potential Antistatic and Slip Additives: a Molecular Dynamics Study}

\author{María del Mar Cammarata$^{1,2}$}
\author{R. Martin Negri$^{1,2}$}
\author{Rocio Semino$^{3}$}
 \email{rocio.semino@sorbonne-universite.fr}
 
\affiliation{1. Instituto de Química Física de Materiales, Ambiente y Energía (INQUIMAE). Consejo Nacional de Investigaciones Científicas y Técnicas (CONICET)-Universidad de Buenos Aires (UBA). Ciudad Universitaria, Pabellón 2, Ciudad Autónoma de Buenos Aires (C1428EGA), Argentina.}
\affiliation{2. Departamento de Química Inorgánica, Analítica y Química Física (DQIAyQF). Facultad de Ciencias Exactas y Naturales. Universidad de Buenos Aires. Ciudad Universitaria, Pabellón 2, Ciudad Autónoma de Buenos Aires (C1428EGA), Argentina.}
\affiliation{3. Sorbonne Université, CNRS, Physico-chimie des Electrolytes et Nanosystèmes Interfaciaux, PHENIX, F-75005 Paris, France.}%

\begin{abstract}

   Additives are essential to enhance or modify the properties of plastics for target applications. However, finding appropriate additives may be challenging, since we lack knowledge on their interactions with plastics and moisture, and the interplay between them. In this work, we study stearoyl diethanolamine (commercial antistatic additive for polyolefins) as well as two amphiphilic molecules as potential new additives for their antistatic or slip properties in polyethylene by means of atomistic molecular dynamics simulations. We reveal that additive/water interactions and relative solubility are strongly determined by their relative ratio. The polyethylene model thin film adopts a crystalline core and an amorphous-like surface, with polymer chain terminations predominantly located at the surface of the slab. Water forms a layer on top of the polymer surface or droplets when its concentration is lowered, but it never enters the polymer matrix. All additives interact with water mainly by their polar heads, with water acting as a hydrogen bond acceptor or donor depending on the additive. The additives studied exhibit remarkably different structures when they are mixed with the polymer: two of them enter the polymer matrix to various degrees, either by intercalating their chains with the polyethylene ones or by forming micellar-like structures, while the third one stays at the surface. When water is incorporated into the system, the structure of some of the additive/polyethylene systems changes. The magnitude and nature of these changes depend on the relative concentrations of all species and on the nature of the additive. The additives penetrate and organize the polymer to some extent, depending on whether water is present. 
   We find that the ethoxylated amine additive orients its polar heads to the surface and its non-polar tail to the center of the film, consistent with its antistatic properties and migratory behaviour. We propose that one of our two modeled molecules could have promising properties as a slip agent, as its behavior in the PE matrix resembles that of the industrial slip agent erucamide. We hope that our predictions will spark interest in testing these molecules in the laboratory as polyethylene additives and in performing similar studies for other additive/polymer pairs.  
\end{abstract}

\maketitle

\section{\label{sec:intro}Introduction}

The masterbatch industry provides polymer formulations in pellet form for the fabrication of final plastic products with a wide variety of properties and target applications\cite{Kosiski2022, Marturano2017, walp2000antistatic, butuc2017antistatic, Ramrez2001, Ramirez2005, coelho2015synthesis, dulal2017slip, dulal2018migration}. The use of additives in these polymer formulations is essential to tailor the properties of the original resin, both within the bulk material and on its surface\cite{ambrogi2017additives}. One of the most widely used polymers in this industry is polyethylene (PE). Many different syntheses and processing methods are currently performed at an industrial level to obtain linear or branched PE with different densities, making PE a tunable polymer for a large number of everyday life applications\cite{coutinho2003polietileno}. Furthermore, its nonpolar character and its low reactivity drive the research and development of compatible additives, considering productive, economic, and environmental aspects\cite{aurisano2021enabling}.

A wide variety of additives for PE is available depending on the polymer of the formulation, such as antioxidants, UV protectors, plasticizers, flame retardants, thermal stabilizers\cite{ambrogi2017additives}, anti-blocking agents, glidants\cite{2014}, and antistatic agents\cite{grossman1993antistatic}, among many others\cite{fink2010concise}. In particular, a central issue for several industries, such as packaging, is that of static electricity on the surfaces of various products. To reduce static in polymers, antistatic additives are used. In polyolefines, compounds from the ethoxylated amine family\cite{butuc2017antistatic, walp2000antistatic} are excellent antistatic agents, as they reduce the surface resistivity of the polymer, favoring the dissipation of electrical charges\cite{Yudaev2023, Kosiski2022,walp2000antistatic, coelho2015synthesis}. Moreover, the material must exhibit proper sliding during processing, such as in extruders. Slip additives are frequently incorporated into the polymer formulation to reduce the friction coefficient and thus improve the sliding ability. For polymers like PE, erucamide is one of the most effective slip agents known up to date\cite{coelho2015synthesis, Ramirez2005, dulal2018migration, 2017}.

Depending on how the additives are incorporated into the polymers, they can be external or internal. The latter ones can be classified as non-migratory or migratory. Migratory additives are commonly amphiphilic molecules, with a polar head that interacts with the external environment (for example, ambient moisture) on the surface of the material. Typically, they must have the ability to migrate from the bulk of the material to its surface, which could be induced by the incompatibility between their polar region and the hydrophobic nature of the polymer matrix\cite{butuc2017antistatic}.
In the case of antistatic additives, it is generally hypothesized that once the polar head is exposed to the external surface, attractive or repulsive interactions are established with ambient water molecules, which are adsorbed on the surface. With respect to ethoxylated amines, it is assumed that attractive interactions with surface water can promote the formation of conduction channels that decrease surface resistivity, $R_s$\cite{grossman1993antistatic, butuc2017antistatic}.

New, more environmentally friendly, cheaper, and more effective migratory additives are needed\cite{Aleksanyan2024, Almeida2022, Kosiski2022}. In a previous study, Cammarata \textit{et al}\cite{MarCammarata2024} experimentally studied the changes in $R_s$ in linear low-density polyethylene (LLDPE) films as a function of the concentration of a series of amphiphilic additives. These compounds contain a sugar-based polar head and an ethoxylated amine, referred to here as EA, whose structure is shown below in Figure 1. In general, the authors observed that the compounds with the greatest reduction in $R_s$ also exhibit a drastic reduction in the water-surface contact angle. In addition, when water droplets slid on a polyethylene surface containing these compounds, a proton transfer from the surface to the water droplets was observed. These studies indicate an important interaction between the three components present at the surface: polymer, additive and water. 

Tailoring these interactions by modifying the chemical groups and concentration of the additives can help tune the surface properties of the polymeric material. Computer simulations have been applied to studying these kinds of systems with promising results. 
Li and coworkers and Wang and colleagues have studied the microstructure and diffusion coefficients of a series of migratory additives in polyethylene by atomistic molecular dynamics simulations, providing insight into the migratory process\cite{Li_2017,Wang_2018}. Cammarata and coworkers performed coarse-graining molecular simulation studies to study the kinetics of the migration process in PE\cite{Cammarata_2023}.
The interactions of water with a bulk PE matrix\cite{Fukuda_1997,Fukuda_1998,Borjesson_2013} and with its surface\cite{Fan_1995,Hirvi_2006,Sethi_2022,Zhang2023}, have also been explored by simulation techniques. However, to the best of our knowledge, no study has yet focused on the interplay between polymer/additive, polymer/water, and water/additive interactions and their impact on the molecular arrangements at the surface of the material for ternary systems. Studying this is crucial to establishing rational design principles for improving plastic formulations. 

In this work, the interactions between a linear and low molecular weight PE, a series of additives, and water at the surface of the material are studied by molecular dynamics simulation techniques. Two organic amphiphilic molecules, referred to as compounds A and B (see Figure 1), were explored as prospective additives for PE. These compounds were chosen because their chemical structures are similar to two compounds evaluated in a previous work\cite{mythesis}, but with smaller polar heads. In this previous study, additives formed by a hydrocarbon chain bonded to a sugar moiety through a 1,2,3-triazole group were experimentally tested for their antistatic and slip properties. Results have shown that one of them strongly reduced surface resistivity when applied over polyethylene films\cite{mythesis}, thus suggesting a promising antistatic agent character. From those results, it follows that analogue molecules with smaller polar heads (i.e., without the 1,2,3 triazole group, see additives A and B in Figure 1) might also exhibit promising antistatic or slip agent behavior, with more effective migratory mechanisms. To compare the behavior of these additives with a reference example and validate the model, an ethoxylated amine, referred to as EA, was also investigated herein. The interplay of polymer/additive, polymer/water, and water/additive interactions as a function of composition was revealed, and the molecular arrangements that result from them were rationalized. The orientations of the additives at the polymer surface are particularly investigated, as this is crucial for their effective activity. This work offers unprecedented insight into the structural complexity of water/polyethylene/additives ternary and binary systems, and the impact of the composition on it. Moreover, we found that additive B exhibited potential slip-agent behavior, suggesting that further experimental evaluation of its performance in polyethylene matrices would be highly promising. We hope that our predictions will motivate the exploration of these additives for the PE packaging industry and that our work will serve as a roadmap for further simulation work on the topic.  

\section{Methods}

All molecular dynamics simulations (MD) were performed with the LAMMPS package\cite{LAMMPS}. Binary and ternary mixtures containing polymer (with water, additives, or both) consisted of a polymer slab of about 30 \AA \ depth, to model a linear and high-density PE thin film containing additives and environmental moisture. Given that the polyethylene/vacuum interfacial region alone spans roughly 20 \AA, \mbox{\cite{Sgouros2017}} a film of this thickness is dominated by interfacial phenomena and disjoining pressure effects;\mbox{\cite{Peng2015}} thereby precluding a meaningful bulk-density analysis.

For each system, total, potential and kinetic energy were collected and plotted to decide whether equilibrium was reached. Intermolecular interactions were studied by analyzing the radial pair distribution functions (RDFs) between relevant atom types. The VMD software\cite{HUMP96} was used to compute the RDFs within the production runs and to visualize the trajectories. 3D periodic boundary conditions were considered in all calculations. In the RDFs of the systems with PE, the values remain above unity at large distances because radial distribution functions were normalized using the total simulation box volume rather than the effective slab volume.

\subsection{Systems and simulation conditions}

Pure polyethylene, water and additives were first simulated at the NPT ensemble (T = 300 K and P = 1 bar) to assess the quality of the individual force fields in terms of the equilibrium density.

Binary systems were divided into three groups: i) water-additives, ii) PE-water, and iii) PE-additives to study the intermolecular interactions between pairs of components independently. In the first group, two cases were considered: systems with excess water  \textbf{(additives in water)}, which consisted of 6 or 7 additive molecules in a 35 \AA\ side box with 908 water molecules, and systems with excess additive \textbf{(water in additives)}, formed by 7 water molecules and 35 additive molecules in a 30 \AA\ side box. These amounts were calculated to keep solvent:solute volume ratio constant in both systems. For the PE-water system (group ii), 211 water molecules were added to the polyethylene slab to obtain the same volume of PE and water in the simulation box. With the same premise, PE-additive systems (group iii) were built by adding 9 additive molecules to the polymer slab.

Two kinds of ternary systems were simulated: \textbf{PE with additives in water, called herein PE-additive-excess water}, and \textbf{PE with water in additives, referred to as PE-water-excess additive}. For the systems with excess water, 4 or 5 additives (depending of the additive studied) and two different water molecules amounts were added: 220 and 25, in order to asses the effect of the amount of water in the system. In the systems with excess additive, 5 water and 8 additive molecules were added.

Five independent replicas were generated for each system to guarantee statistical significance of the results by varying initial positions and velocities during simulation box construction. All replicas were equilibrated and their production runs analyzed. 

The initial state of all the binary and ternary systems was built by randomly adding the extra components (water, additives, and their combinations) over the PE slab via the LAMMPS \emph{create\_atoms} command\cite{lammpsCreate_atomsCommand}. The systems were then left to evolve during the equilibration run. The length of the equilibration runs was tailored to guarantee that each system had achieved equilibrium. All the production MDs were performed at 300 K with a 0.5 femtosecond timestep and lasted between 5 and 10 ns. 
Systems without polymer were run in the NPT ensemble at 1 bar, whereas all the systems that had a polymer slab and thus needed one part of the box to be void, were simulated in the NVT ensemble. Temperature and pressure were regulated using the Nosé-Hoover thermostat and barostat, with relaxation timesteps of 0.3 ps and 0.5 ps, respectively. 

\subsection{Polymer generation}

PE was modelled with the General Amber Force Field (GAFF)\cite{wang2004development}, which consists of harmonic potentials to model bonds and angles, cosine-based potentials for dihedrals and a summation of 12-6 Lennard Jones (LJ) and Coulomb potentials for non-bonded interactions. The LJ and Coulomb short-range cutoff was set to 12 \AA. The long-range part of electrostatics was treated with the particle-particle particle-mesh solver, with a relative error in the forces of 10$^{-5}$. Partial atomic charges were assigned after building the polymer, as explained in the following paragraphs. 

PE was constructed \textit{in silico} via the methodology developed by Abbot \textit{et al}\cite{abbott2013polymatic} as implemented in the \textit{Polymatic} code\cite{psuColinaGroup}. This building strategy was selected because it is appropriate for modelling amorphous, linear, and polydisperse polymers, a fitting description for high-density polyethylene. In addition, it is known that migration of amphiphilic additives takes place mainly in the amorphous regions of the polymers\cite{fang2017predicting} rather than the crystalline ones, so modelling an amorphous polymer should be advantageous for studying potential migratory additives. The polymer construction procedure comprised 5 stages:

\textbf{1)} Monomer description: the monomer consisted of a -\ce{CH2CH2}- moiety, so that the two carbon atoms are undercoordinated. Five atom types were initially defined: three carbon and two hydrogen types. One of the carbon types was described as "reactive", which means that it can form C--C bonds. Reactive carbons become regular carbon atoms after polymerization. The "reactive" carbon type is thus absent in the final polymer model. The third carbon atom type is reserved for polymer terminations, and it is assigned to the chain ends at the end of the polymerization process. Initial partial atomic charges were assigned to each atom from a previously reported\cite{marenich2012charge} PE model: $q_H=$0.079 and $q_C=$-0.236 (in elementary charge units). As our monomer contains fewer hydrogen atoms, the resulting excess negative charge was compensated by increasing $q_H$ to 0.118 to ensure overall electrical neutrality. These values were subsequently readjusted after polymer construction, as discussed below.  

\textbf{2)} Simulation box building: A cubic, 50 \AA\ side box was filled with 1290 monomers, which corresponds to an initial density of 0.48 g/ml. Commercial high-density polyethylenes have a density in the range of 0.94-0.97 g/ml\cite{coutinho2003polietileno, Jung2018}, but a lower initial density is recommended to facilitate the polymerization with \textit{Polymatic}\cite{abbott2013polymatic}.

\textbf{3)} Polymerization: three parameters must be set according to \textit{Polymatic} to start polymerization: temperature (T), minimum distance at which two monomers must be from each other to form a bond $(x_{min})$, and number of bonds to be formed per cycle (B). Several polymerization runs were performed until the optimal values for these parameters to obtain the longest possible chains were identified: $x_{min}$ = 10 \AA, T = 500 K, and B = 14. The longest chain had 48 monomers (which means 96 carbon atoms), with the remaining chains containing from 2 to 38 monomers, and a small fraction of unbound monomers.  

\textbf{4)} Adding chain-ends and chains selection: the elongated chains have no terminal hydrogens, as the monomer contains only four hydrogen atoms. The terminal hydrogen atoms were added via an in-house code. After that, the chains selection was made. For this purpose, a final cubic box size of 30 \AA\ was estimated, so that 557 monomers were needed to yield a 0.95 g/ml density (considering the density range mentioned above). Therefore, the longest chains were selected and the number of chains was regulated so that the total sum of monomers was close to 557. This procedure led to selecting chains from 96 to 30 carbon atoms. Shorter chains and unbound monomers were removed from the simulation box using another in-house code. In this way, a non-branched polyethylene model was constructed. The weight-average molecular weight ($M_w$), the number-average molecular weight ($M_n$), and the polydispersity index were calculated according to their definitions\cite{Odian2004}.

At this stage, partial atomic charges were adjusted to account for the addition of terminal hydrogen atoms. To this end, the total number of H and C atoms in the polymer slab were first calculated. Using the previous values of $q_H=$0.118 and $q_C=$-0.236, an excess positive charge was obtained. To restore charge neutrality, this excess was redistributed while maintaining physically reasonable values for each atom. This procedure resulted in values of $q_H$=0.10 and $q_C=$-0.2044 (in elementary charge units).

\textbf{5)} Equilibration and slab building:  Equilibrating polymers is difficult due to the presence of degrees of freedom with very slow dynamics given by the large molecular weight of the polymer chains and their entanglement. This is why the equilibration requires a series of MD simulations and considerable computing time. Therefore, the equilibration was done through an initial MD run in the NVT ensemble, followed by the ``21 steps'' sequence developed by Abbot and coworkers,\mbox{\cite{abbott2013polymatic}} which involved 21 MD simulations, consisting of 7 cycles of two NVT runs at different temperatures and an NPT run at different pressures. This procedure has been effectively used for equilibrating a series of polymers.\mbox{\cite{abbott2013polymatic,Semino2015,Semino2018}} 

A box with a slab unwrapped in one direction ($z$ direction herein) needed to be built, including void space to ensure the periodic images do not interact with each other. To have a dense slab, after unwrapping the polymer, a spring force was applied to the center of mass of each of the chains to tether them to a fixed position, using the \emph{spring force command} available in LAMMPS\cite{lammpsSpringCommand}.  
Then, the slab was further equilibrated through a series of NVT and NPT cycles. 

\textbf{6)} Production runs: Finally, five 10-nanosecond-long independent production MDs in the NVT ensemble (T = 300 K) with a 0.5 femtosecond timestep were performed to obtain characteristic information of the PE model. Results are the average of those obtained for the five replicas.

\subsection{Additives and water models}

\begin{figure}[H]
 	\includegraphics[scale=0.4]{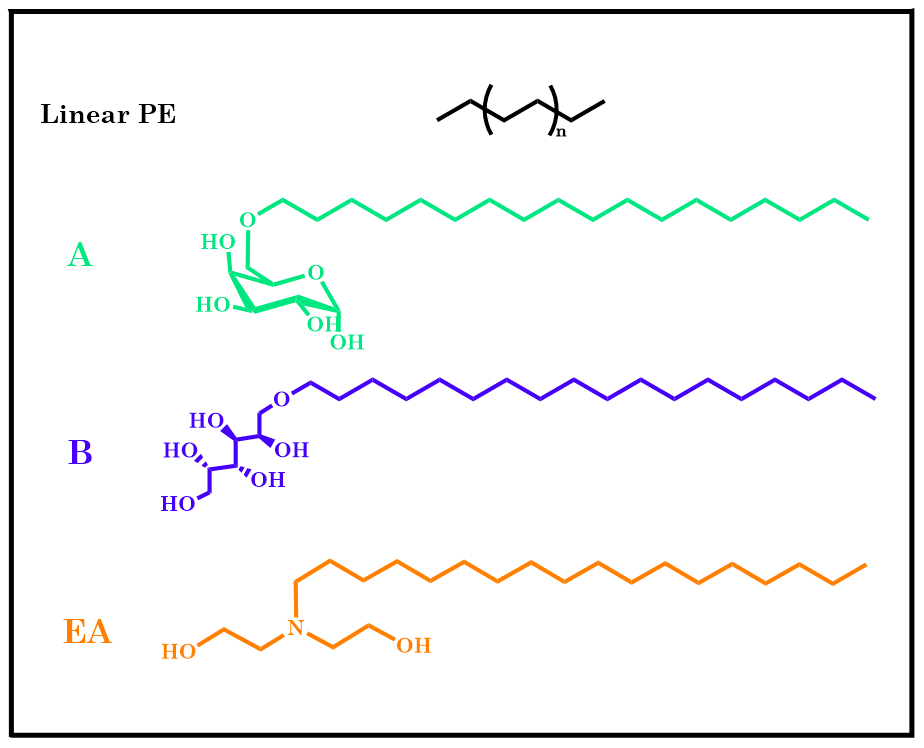}
	\centering
    	 \caption[] {Scheme of the molecules modeled: linear PE and additives A, B, and EA.}  
	\label{fig1}
 \end{figure}

Simulated additive molecules are shown in Figure \ref{fig1}. Each additive has a distinctive colour, which will be maintained throughout the text: green for additive A, violet for B and orange for EA. 

The GAFF-derived force field GLYCAM06\cite{kirschner2008glycam06}, suitable for carbohydrates, was employed for the additives parametrization. All potential functional forms and cutoffs are those described above for PE. In GLYCAM06, sugars like galactose have carbon atoms with a partial positive atomic charge due to the electrowithdrawing effect of oxygen atoms. In contrast, carbon atoms in a non-polar hydrocarbon chain have a negative charge, as seen in polyethylene. Hence, two different carbon atom types must be defined to parameterize all additives. Partial atomic charges defined in GLYCAM06 were used for the sugar-moiety carbon atoms in additives A and B, and the same atom types defined for the polymer were used to model the carbon atoms in the hydrocarbon chains. Slight adjustments were made when necessary to ensure electroneutrality. Figure S1 shows the structures of the simulated molecules labeled according to their atom types. Partial atomic charges of EA were based on the monoethanolamine force field developed by Alejandre and coworkers\cite{Alejandre2000}. 

A simple, flexible force field was selected for water to achieve a more realistic description of its interactions with PE and the additives. Flexibility allows to model water vibrations that may affect hydrogen-bonding interactions. The parameters of this force field can be found in the Supporting Information (SI)\cite{jorgensen1983comparison}. 

\section{Results and Discussion}
\subsection{Pure systems}

The PE model built contained 24 linear chains with different lengths and molecular weights (see SI, Table SIV). The values obtained for $M_w$ and $M_n$ were 716 kDa and 642 kDa, respectively, with a polydispersity index of 1.1. 

That initial polymer slab was amorphous, but after the production dynamics, the chains acquired a partially ordered structure in all cases while maintaining an amorphous region at the top and bottom of the box in the z-direction, as shown in Figure S2a,b. This ordering is attributed to two main factors. First, the short (less than 100 C atoms), non-branched chains in our PE model may be more prone to form crystalline-like domains than longer chains, which could form entanglements that locate mainly in amorphous domains, as demonstrated by Sliozberg and collaborators\cite{Sliozberg2018}. Moreover, as the PE monomer contains two carbon atoms, all the chains in the final polymer have an even number of carbon atoms. Thus, they are able to adopt extended linear forms, as observed by Prisacariu and Scortanu in a previous study\cite{Prisacariu2011}. Extended linear forms favor a more effective packing between chains, resulting in the presence of crystalline domains in the simulated polymer. 

It is worth noting that the methyl chain ends (highlighted in purple for carbon atoms and yellow for hydrogen atoms in Figure S2a, b) are mostly located at the surface of the polymer slab rather than in the ordered core of the polymer (see the density profile for polymer terminations in Figure S3). This observation is in agreement with the results of Lee \textit{et al},\cite{Lee2020} who studied a similar system of linear but longer PE chains (up to 150 carbon atoms). They found a surface segregation of methyl chain ends that is more likely to take place in long PE chains, and more pronounced when the temperature decreases, with 300 K the lowest temperature explored (which matches the temperature used here). The authors concluded that this surface segregation arises from the enthalpic origin attributed to the lower cohesive energy density of the terminal methyl groups. Furthermore, they also found a segregation of shorter chains to the surface in both the moderately and highly asymmetric mixtures of two different chain lenghts\cite{Lee2020}. This was not observed in our system due to the presence of the tethering potential, which constrains the chains to remain localized at fixed positions along the $z$ axis. Lakkas and coworkers also reported the accumulation of chain terminations at the surface of polymer slabs via a combination of self-consistent field theory and atomistic simulations.\mbox{\cite{Lakkas2019}}

Additionally, the ordering might also be related to the fact that the simulated temperature (300 K) is above the vitrimer transition temperature of the polymer, which is experimentally found in the range 190-200 K in amorphous polyethylene\cite{shalaby1981thermoplastic}. This confers the polymer a higher mobility that allows it to adopt more ordered configurations. 

The average C--C RDF obtained from the five independent runs of the PE system (Figure S2c) clearly shows the partially crystalline structure, judging by the multiple peaks at distances above 3 \AA. The result agrees with those obtained by Narten for a PE powder with $M_w$ = 90000 g/mol at 300 K\cite{Narten1989}, indicating that the chosen force field and the methodology employed properly capture PE structure. 

An average density of 1.00 g/ml was obtained for pure water with the employed flexible model. This is an expected value in agreement with experimental data.   

Additives A and B exhibited 0.98 g/ml and 0.99 g/ml densities, respectively. No experimental data on these chemical compounds are available. Still, the density values are close to those predicted by the tools integrated in $SciFinder^n$ CAS\cite{casSciFinderxAELogin,casSciFinderxAELogin2} for each molecule: $1.06 \pm 0.06$ g/ml for additive A, and $1.04 \pm 0.06$ g/ml for additive B. These values are obtained through QSPR/QSPAR models for physicochemical properties prediction\cite{Cherkasov2014, Todeschini2009}.
A 0.90 g/ml density was obtained for EA at 300 K, in good agreement with the value of 0.88 g/ml reported at 323 K\cite{komori1956}.

RDFs analyses of additive A revealed peaks at 1.95 \AA\ and 3.55 \AA\ for the $O_{h}-H_{o}$ pair (see Figure S1), corresponding to intramolecular distances in the galactose moiety with different conformations of the pyranose form\cite{arabal2005}. Peaks at 1.99 \AA\ and 3.56 \AA \ between the same atom types were obtained for additive B, consistent with reported $O_{h}-H_{o}$ distances in alcohol clusters\cite{Provencal2000}. The $O_{h}-H_{o}$ bonding distance was between 0.95 \AA\ and 0.97 \AA\ both for additives A and B, in good agreement with the typically reported O--H bond distance of 0.96 \AA\ to 0.97 \AA\cite{Provencal2000}.

The successful comparison between experimental and predicted data allowed validating the force fields employed for the PE, water and additives.

\subsection{Binary systems}
\subsubsection{Additives and water mixtures}

The interactions of the three additives with water were studied both at excess water and excess additive conditions. Four pairs of atoms (see Figure S1) were analyzed to characterize these interactions through RDFs: \textbf{$C_{g}-O_{w}$} and \textbf{$O_{w}-H_{c}$} to see interactions between water and the less polar fraction of each additive (hydrocarbon chain), and \textbf{$O_{w}-H_{o}$}, \textbf{$O_{h}-H_{w}$}, and $N_{T}-H_{w}$ (only for the case of EA) to account for hydrogen bonds and polar interactions.

Visual inspection of the \textbf{additives in water} trajectories revealed that in all cases the additive molecules remained together during the simulations, forming aggregates in bulk water (see Figure S), which is consistent with their global non-polar character. In the case of additives A and B, the aggregates are arranged with the galactose moieties facing each other, while the hydrocarbon chains extend outward, resembling a lipid bilayer-like organization (Figure S4).
This behavior is less clearly defined in the EA system. Although bilayer-like arrangements are occasionally observed in some EA-in-water trajectories, the molecules are more frequently aligned, with the ethoxylamine groups oriented in the same direction and the hydrocarbon chains packed together on the opposite side, as shown in Figure S4. Overall, the results suggest that the additives are not soluble in water, even when a few additive molecules are present. 

Repulsive interactions were identified between the three additives and water by looking at the $C_{g}-O_{w}$ and $O_{w}-H_{c}$ pairs. Indeed, the interaction frequency is below the bulk reference at short distances, as seen in Figure S5. This is expected due to the non-polar character of hydrocarbon chains. 

\begin{figure}[H]
 	\includegraphics[scale=0.9]{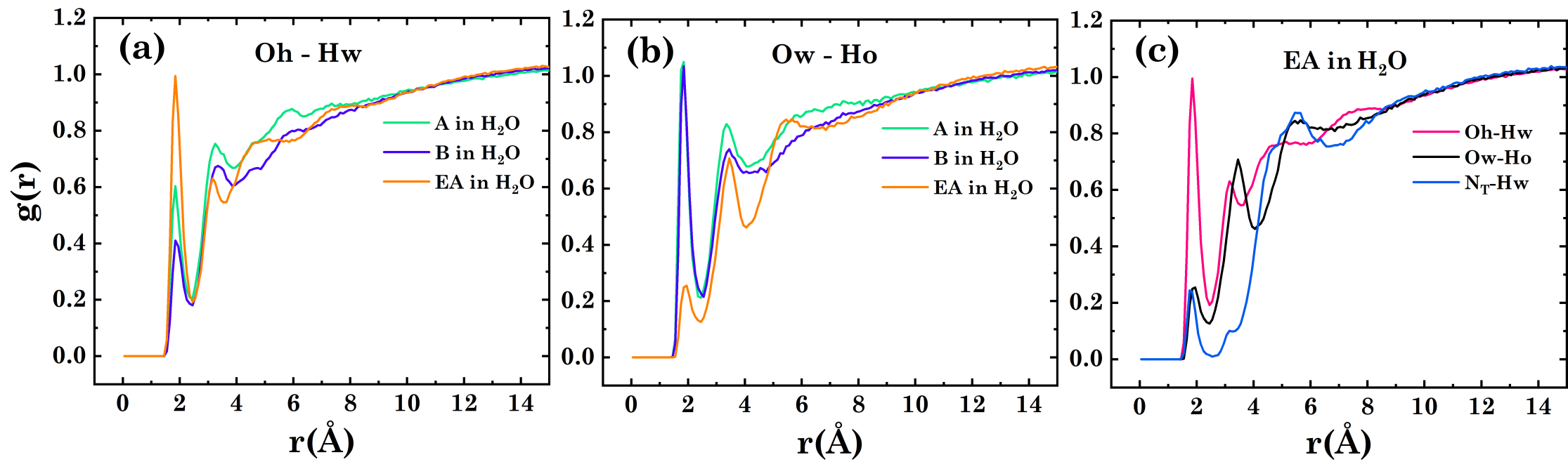}
	\centering
    	 \caption[estructuras de agua] {Average of the radial pair distribution functions of the \textbf{additives in water} systems for \textbf{(a)} $O_{h}-H_{w}$ pair, \textbf{(b)} $O_{w}-H_{o}$ pair. Different colors identify additives, with A in green, B in violet, and EA in orange. \textbf{(c)} Average of the radial pair distribution functions of two possible hydrogen bonds in the EA-in-water system.}  
	\label{hbonds-axenagua}
 \end{figure}

\begin{figure}[H]
 	\includegraphics[scale=1]{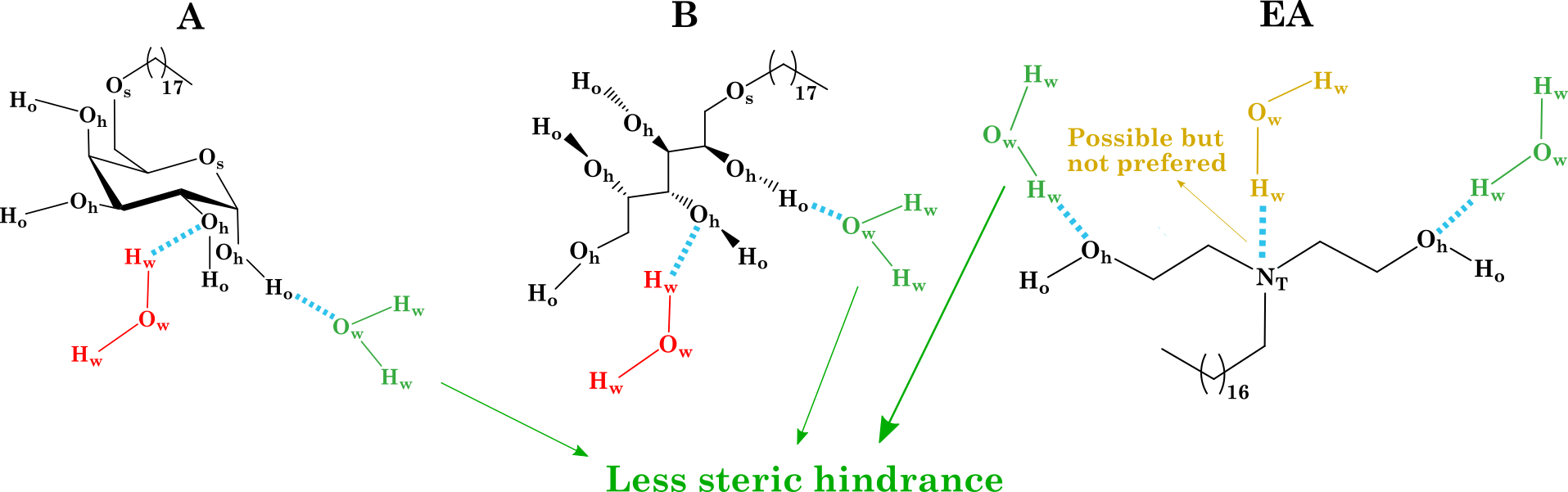}
	\centering
    	 \caption[estructuras de agua] {Additive molecules representations focused on their polar regions. Atom types are labeled as described in the text. $O_{h}-H_{o}$ bonds of hydroxyl groups are explicitly drawn to show the hydrogen bonds established in each case. Hydrogen bonds are represented with light blue dashed lines. Water molecules are drawn in red and green to represent unfavorable and favorable orientations towards the polar groups of additives, respectively. For A and B, sugar moieties generate steric hindrance for the $O_{h}$ atoms. The smaller polar group ethoxyamine of EA allows water to establish hydrogen bonds between $O_{h}$ and $H_{w}$.}  
	\label{sterichindrance}
 \end{figure}

A more interesting result was obtained from the $O_{w}-H_{o}$ and $O_{h}-H_{w}$ (see Figure \mbox{\ref{sterichindrance}} to identify the atom pairs involved), whose RDFs are shown in Figure \ref{hbonds-axenagua}. An important feature of the hydrogen bond interactions in the systems can be extracted from these RDFs: for $O_{h}-H_{w}$, the functions have similar shapes for the three additives, but only for EA (orange curve in Figure \ref{hbonds-axenagua} a) the frequency is close to the bulk value at a distance of 1.85 \AA. The opposite trend is observed in the $O_{w}-H_{o}$ function, where EA presents the lowest frequency at 1.92 \AA \ compared with additives A and B. This indicates hydrogen bonds between water and EA are preferably established between the oxygen of the EA $(O_{h})$ and the hydrogen of water $(H_{w})$. In contrast, for A and B, they are established between the oxygen of the water $(O_{w})$ and the hydrogen of hydroxyl groups of the sugars, $H_{o}$. In the specific case of EA, a third hydrogen bond is possible between $N_{T}$ and $H_{w}$ atoms, but the comparison of RDFs in Figure \ref{hbonds-axenagua}c indicates the $O_{h}-H_{w}$ interaction is favored, as it is observed at a higher frequency than the other two hydrogen bonds and close to the bulk value. 

These results can be rationalized considering that the sugar moieties in A and B are bulkier than the ethoxyamine group. There is a steric hindrance that makes oxygen atoms $O_{h}$ less accessible for $H_{w}$ atoms, and as a result, hydrogen bonds between $O_{w}$ and $H_{o}$ are more favorable for additives A and B. Figure \ref{sterichindrance} illustrates additive structures highlighting their polar regions. Water molecules find less steric hindrance when getting close to the A and B hydroxyl groups by their oxygen atoms. With EA, water molecules can get close through their hydrogen atoms to $O_{h}$ or $N_{T}$ atoms, but $O_{h}$ atoms are preferred because $N_{T}$ atoms are more hindered.

The \textbf{water in additives} trajectories show a similar behavior for additives A and B: water molecules remain dispersed throughout the simulation box and do not exhibit a tendency to aggregate (see Figure S6). They interact with both the hydrocarbon chains and the sugar moieties. For the case of EA, in Figure S6c, a slightly closer proximity between some water molecules can be observed compared to systems with A and B. This is consistent with the RDFs for $C_{g}$--$O_{w}$ and $O_{w}$--$H_{c}$ pairs, illustrated in Figure S7. Interactions between $C_{g}$ and $O_{w}$ are attractive in systems with additives A and B, while they are only weakly attractive in the case of EA, suggesting a lower affinity of water molecules for the hydrocarbon chains in the latter system.

Interactions between $H_{c}$ and $O_{w}$ were repulsive for the three additives, with a more defined peak only for additive B. Both RDFs are consistent with the dispersion of water molecules in the box with additives A and B, allowing water to interact with the polar and non-polar regions of the two additives. In contrast, in EA environments, water molecules interact mainly with the polar groups.

Concerning hydrogen bonds, the same preference for $O_{h}$--$H_{w}$ over $O_{w}$--$H_{o}$ in EA was observed compared to A and B, as shown in Figures S7c and d. 

In summary, under conditions of excess additives, the $C_{g}$--$O_{w}$ interactions are attractive for all three additives (though less intense for EA), whereas $H_{c}$--$O_{w}$ interactions are repulsive in all cases. This suggests that water oxygen atoms preferentially interact with the $C_{g}$ of the additives rather than with $H_{c}$ atoms. This also indicates that water molecules are dispersed within the additives matrix, as previously discussed. Conversely, with excess water, both $C_{g}$--$O_{w}$ and $H_{c}$--$O_{w}$ interactions become repulsive for the three additives, consistent with their tendency to agglomerate. This leads to stronger self-interactions between additive molecules, which in turn repel water.  

\subsubsection{PE with water}

Simulating a realistic scenario where the interface of industrial PE films is exposed to a relative humidity of around 50\% at 300 K would imply adding less than one water molecule in the simulation box. If this system were built, a slab with a much larger surface area would be required. This would result in a much larger one that would be prohibitively expensive to treat at the atomistic level. Hence, the pure PE system might be considered representative to that with a relative humidity of around 50\% at 300 K. Therefore, the interactions studied in the binary system described herein are representative of a PE/liquid water interface. 211 water molecules are placed over the PE slab, representing a layer of liquid water. 

\begin{figure}[H]
 	\includegraphics[scale=0.8]{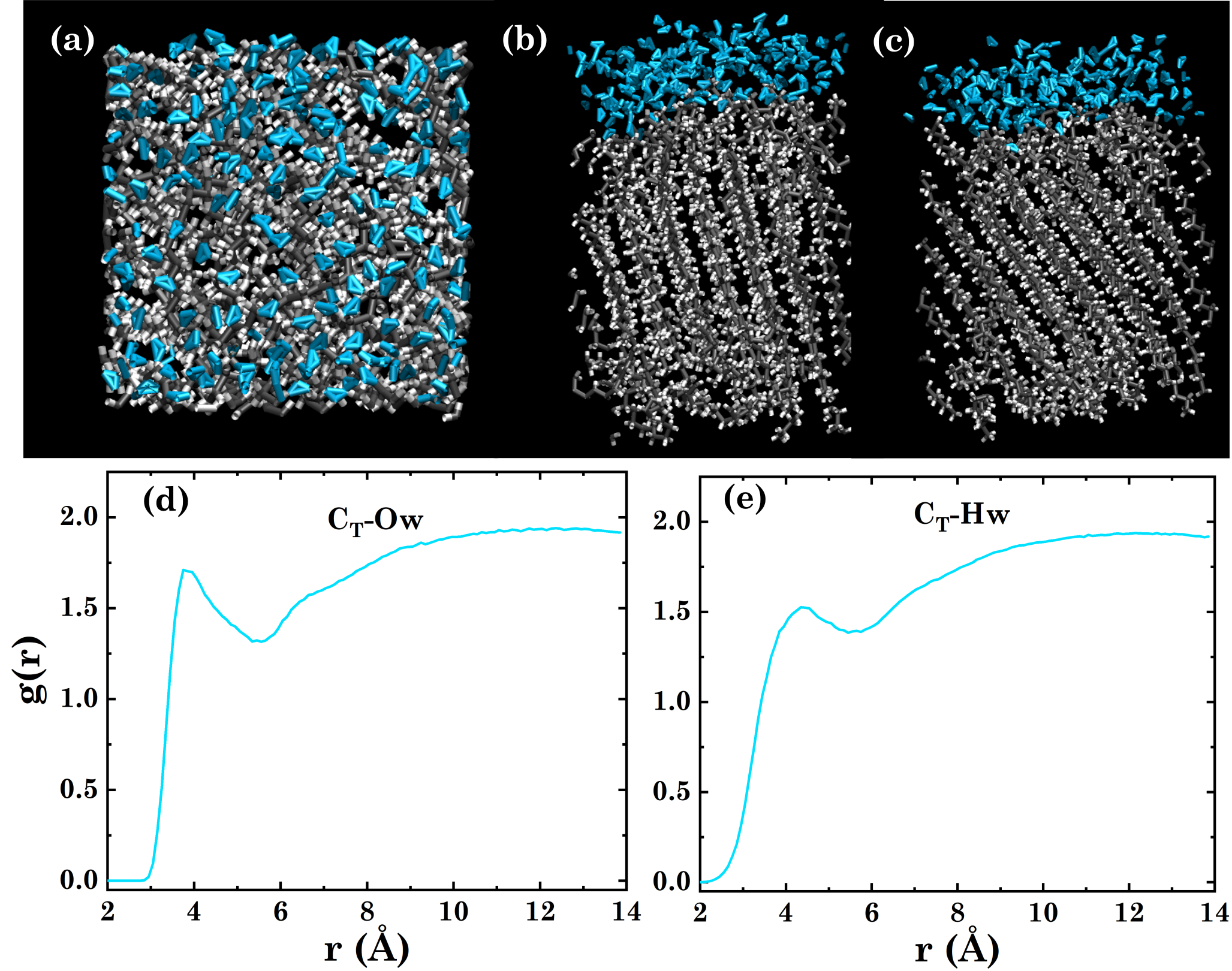}
	\centering
    	 \caption[] {Results for PE-water. \textbf{(a), (b)} and \textbf{(c)}: Snapshots of different runs of the system from different planes; and averaged RDFs (g(r)) of \textbf{(d)} $C_{T}$--$O_{w}$ and \textbf{(e)} $C_{T}$--$H_{w}$ pairs. Peaks below the bulk value indicate weak interactions between terminal carbon atoms of the polymer and water hydrogen and oxygen atoms.}  
	\label{pe-water}
 \end{figure}

PE also acquired the partially ordered structure as previously discussed in the presence of water, Figures \ref{pe-water}b and c show this in two representative snapshots of the simulation box from two different trajectories. The water molecules homogeneously cover the PE-water interface, as shown in Figure \ref{pe-water}a, but remain located over the PE interface, without penetrating the matrix. Due to this, the RDFs for internal PE atoms C and H, with water atoms $O_{w}$ and $H_{w}$ (not shown), exhibited repulsive interactions, as expected. Weak interactions were found only between terminal carbon atoms ($C_{T}$) and water oxygen and hydrogen atoms, as seen in the RDFs shown in Figure \ref{pe-water}d and e, respectively, as these terminal atoms are more abundant at the surface of the polymer slab. The latter observation reinforces the one made before when analysing the pure PE system: the methyl chain ends of the polymer are segregated on the surface, in agreement with the results reported by Lee and collaborators\cite{Lee2020}.

The peaks are close but still below the bulk value of the function, thus indicating a poor interaction at the interface. There is a subtle difference between the $C_{T}$--$O_{w}$ and $C_{T}$--$H_{w}$ RDFs, the former seems slightly stronger. This might suggest a predominant orientation of water molecules with oxygen atoms toward the polymer interface, like in hydrogen-bonding-capable polymers\cite{WalkerGibbons2022}. However, no orientational preference was detected upon further analysis of the trajectory of the PE-water system. Hence, the difference between the RDFs does not seem to be correlated with a significant difference in orientation. If there were a preferred orientation, it would be expected to be with one of the hydrogen atoms near the polymer, as Zhang \textit{et al.} suggested in previous work \cite{Zhang2023} for non-hydrogen-bonding polymers like PE. Instead, the OH group orients normally to the PE surface; this matches the so-called "dangling" OH group\cite{Lee1984}, observed in other hydrophobic surfaces as well, such as alkanes and polydimethylsiloxane\cite{Strazdaite2015}. This orientation allows water molecules to optimize hydrogen bonds at the bulk and minimize the disturbance produced by the non-polar material.

\subsubsection{PE with additives}

\begin{figure}[H]
 	\includegraphics[scale=0.93]{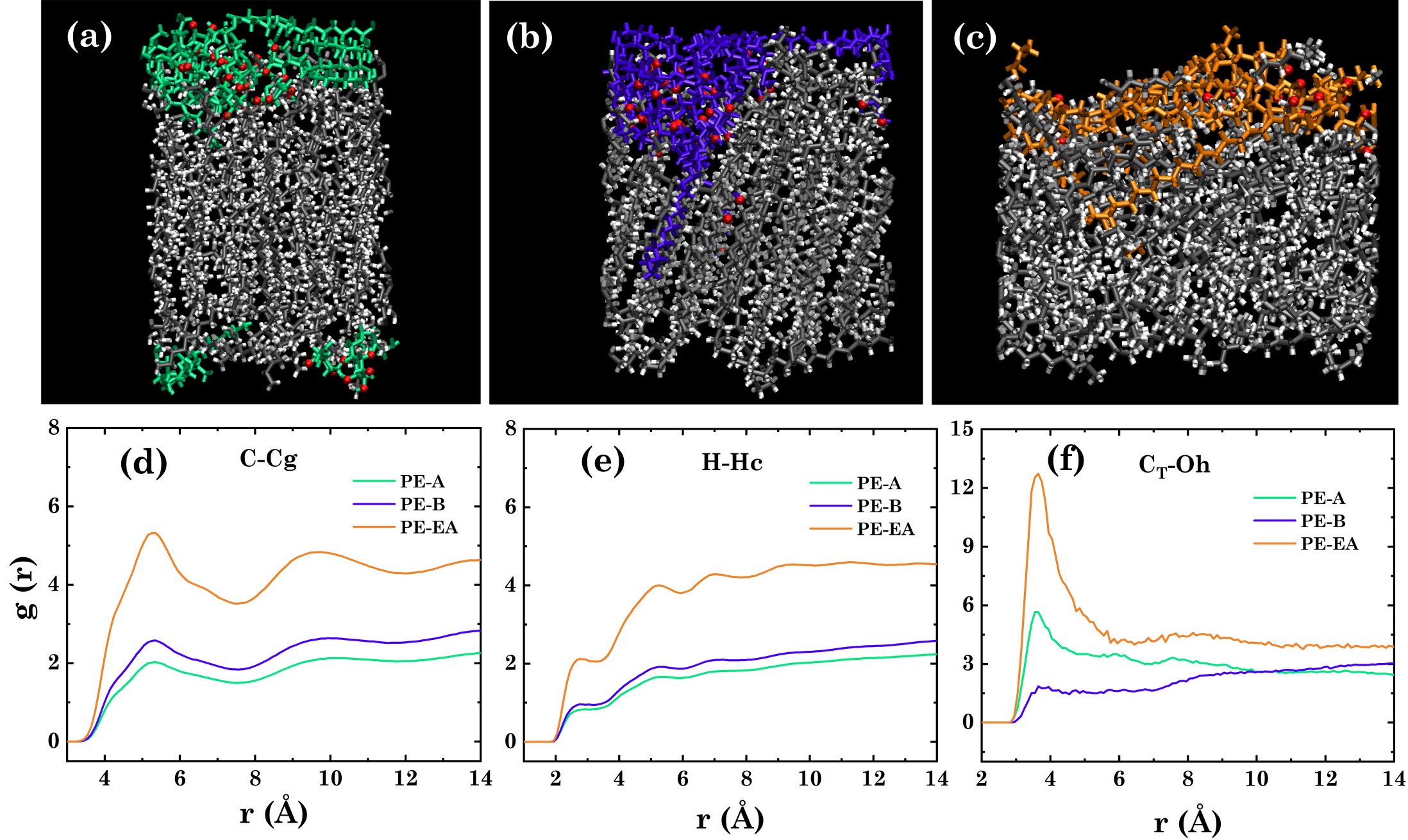}
	\centering
    	 \caption[] {Representative snapshots of PE-additive systems: \textbf{(a)} with additive A, \textbf{(b)} B and \textbf{(c)} EA. \textbf{(d), (e)} and \textbf{(f)}: $C$--$C_{g}$, $H$--$H_{c}$, and $C_{T}$--$O_{h}$ Averaged RDFs for the A, B and EA-containing systems, respectively.}  
	\label{pe-ax}
 \end{figure}

PE with additive systems consisted of the polymer slab with a layer of additives above it and an empty space to avoid interaction of slabs through periodic boundaries. Figure \ref{pe-ax} shows a representative snapshot for each system. The additives are represented with the colors used in Figure \ref{fig1}, and only $O_{h}$ atoms are distinguished as red balls to make it easier to identify the polar regions of the molecules. 

Additive A mostly keeps its initial position over the PE slab surface, occasionally penetrating, while additive B goes deeper into the matrix, with its polar regions (sugar moieties containing red $O_{h}$ atoms) clustering together, forming a polar pocket. This orientation resembles a micelle or a bilayer, which is consistent with the reported ability of 6-O-alkyl-galactitols to form liquid crystals\cite{BAULT1998}. EA additive molecules are concentrated near the surface, but more immersed than additives A and B, in a closer interaction with the PE matrix. 

These results can be interpreted by considering two combined effects: the relative polarity of the sugar and ethoxyamine moieties, and the affinity between the hydrocarbonated chain of the additives and the PE matrix. Regarding the latter, RDFs in Figure \ref{pe-ax}d-f reflect a higher affinity of EA with PE: $C$--$C_{g}$ interactions are attractive for EA, but repulsive for A and B, as these additives remain mostly at the PE surface. Although the $H$--$H_{c}$ RDFs show a repulsive interaction for the three additives, the more defined multiple peaks indicate a weak long-range interaction between those atom types in the PE-EA system compared with additives A and B, thus suggesting a higher affinity of EA for PE. 

Lastly, for the $C_{T}$--$O_{h}$ pair, a strong attractive interaction is seen for EA, while it is weaker for A, and repulsive for B. Terminal carbon atoms of PE $(C_{T})$ are closer to galactose moieties and ethoxyamine groups because they are near the interface. In contrast, additive B adopts a different arrangement: polar groups (galactitol moieties) cluster together inside the PE core instead of away from it, at the surface. The previously mentioned bilayer-like configuration for additive B was noted for the commercial slip agent erucamide\cite{dulal2018migration, coelho2015synthesis}, as reported by Dulal \textit{et al}\cite{dulal2017slip}. This similarity makes additive B interesting to be further studied as a potential slip agent for polyethylene. 

It is useful to consider the potential presence of water to better assess the influence of the relative polarity of the sugar and ethoxyamine moieties on the properties of the additives. As discussed for the case of the PE-water system, simulating 50\% humidity at 300 K would involve adding less than one water molecule in the simulation box. Therefore, the binary PE-additive systems can be analysed as the closest scenario to that with low water content. From this point of view, the different behavior observed in the snapshots shown in Figure \ref{pe-ax} could be related to a different affinity of the polar moieties of each additive with water. While the ethoxyamine groups of EA are at the interface of PE, widely available for interacting with the environmental moisture, the galactose moieties of additive A seem to be less available, and the galactitol groups of additive C are clustered together. This allows us to establish a clear trend of polar moiety-water affinity, which decreases from EA to additives A and B. This trend is consistent when compared with experimental data on the water solubility of the molecules resulting from removing the 18-carbon chain from each additive. This narrows the analysis to a direct comparison between galactose, galactitol, and diethanolamine. A very high solubility in water is reported for diethanolamine (1 Kg/L), which is equivalent to a concentration of 9.5 M\cite{Dow1980Alkanolamines}. Moreover, as this compound is liquid at room temperature, it is totally miscible with water, or simply water soluble\cite{CAMEO_EthyleneCarbonate}. The reported solubility in water of galactose is 2.25 mol/L at room temperature (close to 298 K)\cite{Yalkowsky2010}, which is 4.2 times less soluble than diethanolamine. Finally, galactitol achieves a similar solubility to galactose of 2.08 mol/L\cite{Yalkowsky2010}, only at 100 °C, which means its solubility is even lower at room temperature. This fact is also related to the symmetry of galactitol, given by the stereochemistry of each carbon atom, which allows it to crystallize efficiently\cite{Berman1968}, resulting in a higher melting point than $\alpha$-D-galactose (188 °C\cite{Berman1968} compared to 169°C\cite{Yalkowsky2010}, respectively), and a clearly lower solubility in water when compared with its analogue compound sorbitol\cite{Yalkowsky2010}. 

In summary, the affinity of each polar moiety with water seems to control the disposition of additive molecules in the binary systems with PE, as seen in Figure \ref{pe-ax}.

\subsection{Ternary systems}

\begin{figure}[H]
 	\includegraphics[scale=0.8]{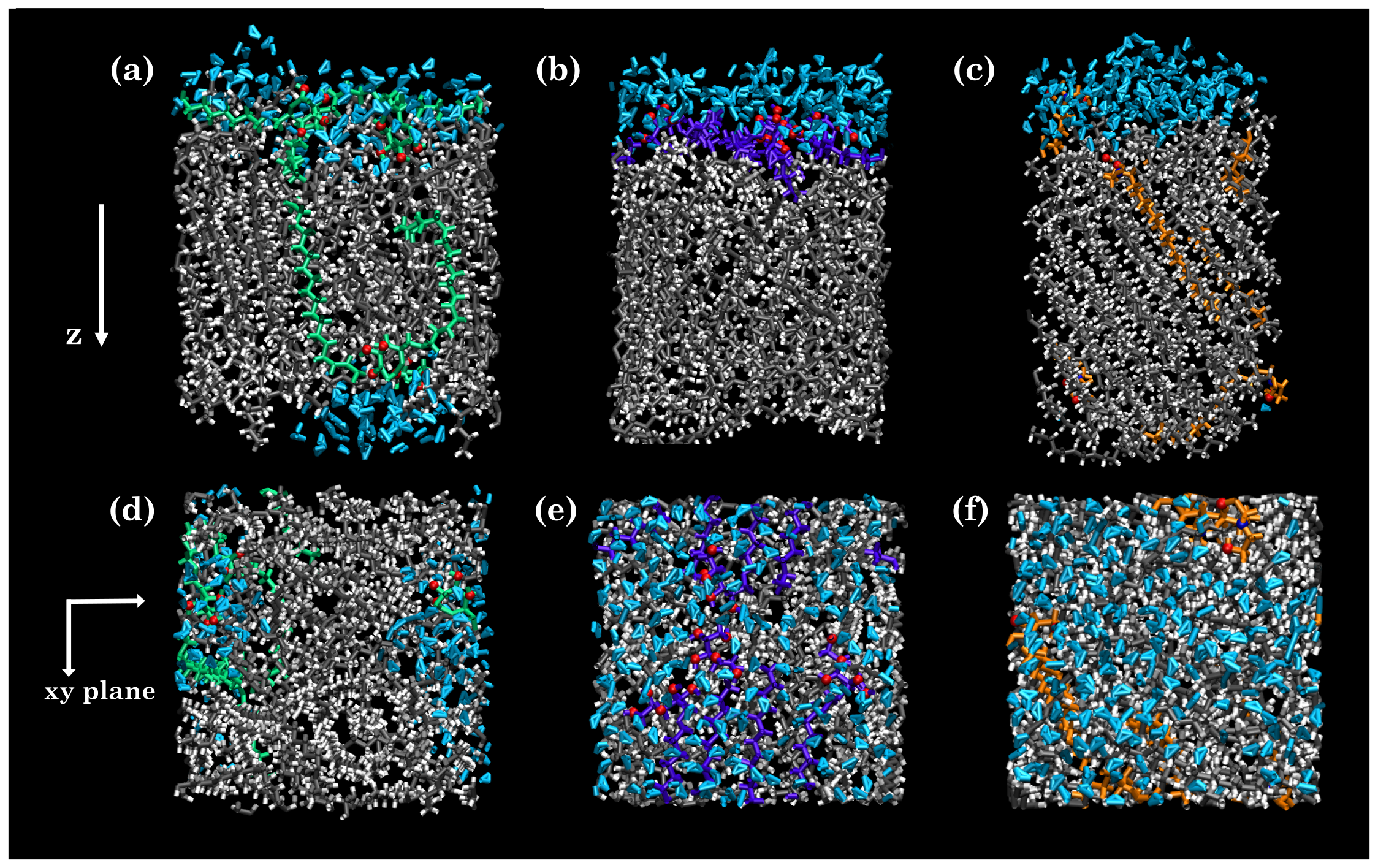}
	\centering
    	 \caption[] {Snapshots of the PE-additive-excess water systems: \textbf{(a)} containing additive A, \textbf{(b)} B and \textbf{(c)} EA. \textbf{(d), (e)} and \textbf{(f)}: snapshots of the xy plane view for the same systems, respectively.}  
	\label{pe-axinwater}
 \end{figure}

Figure \ref{pe-axinwater} shows the final frame of a representative \textbf{PE-additive-excess water} trajectory for each system, with the same color codes as in Figure \ref{pe-ax}, from two different planes, as indicated in the figure caption.

\begin{figure}[H]
 	\includegraphics[scale=0.75]{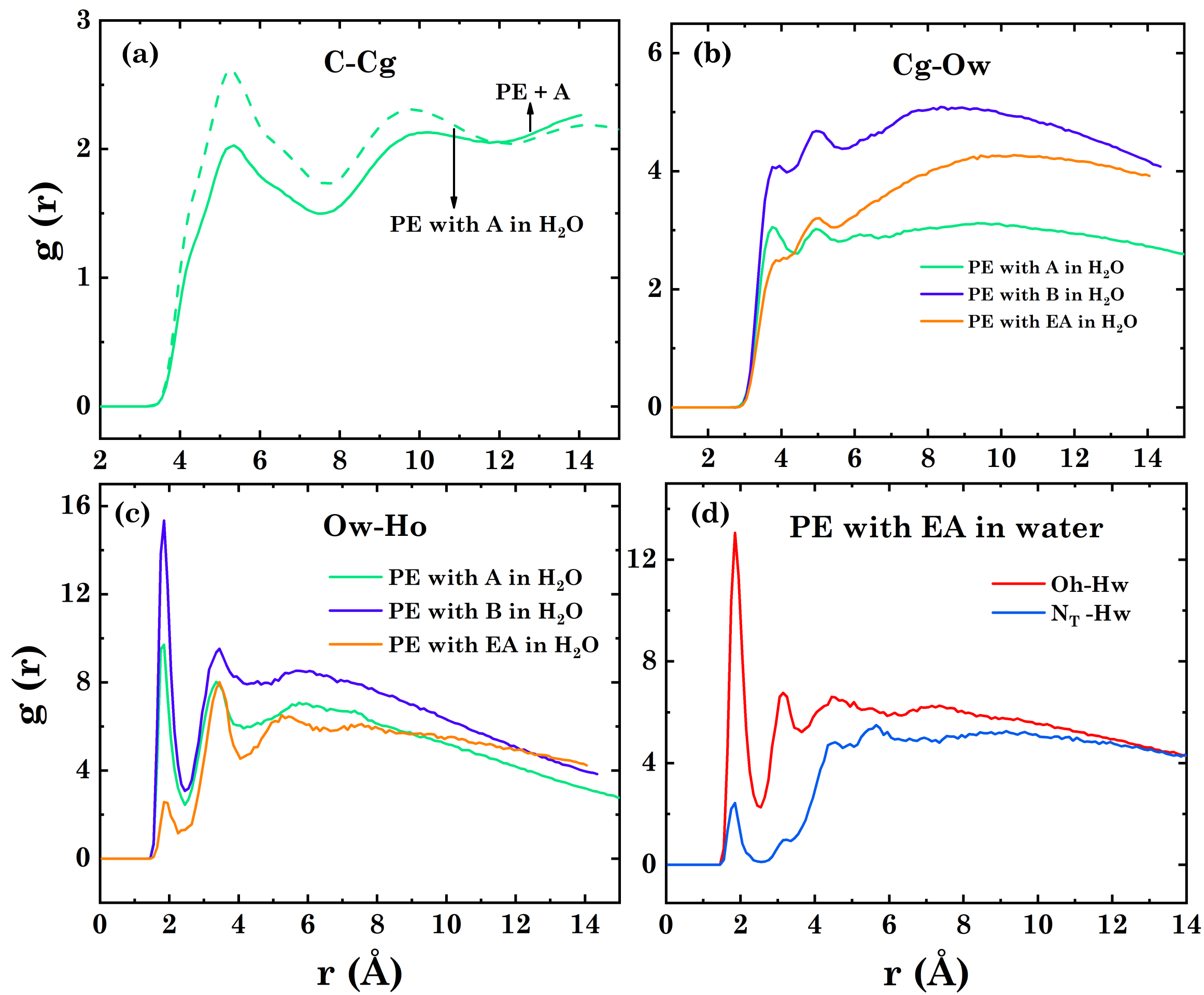}
	\centering
    	 \caption[] {Selected RDFs from PE-additives-excess water systems.  Comparison of RDFs for the PE-additive A binary system and its analogous with excess water; \textbf{(a)} $C$--$C_{g}$ \textbf{(b)} $C_{g}$--$O_{w}$; \textbf{(c)} $O_{w}$--$H_{o}$, and \textbf{(d)} Comparison of RDFs for two of the three possible hydrogen bonds between EA and water.}  
	\label{ternarios7}
 \end{figure}

When water molecules are located over the polymer, additive A chains exhibit a strong repulsion with water, which leads to their immersion into the polymer, with the hydroxyl groups of the galactose head pointing towards the water layer. This behavior is notoriously different from that of the analog system without water (see Figure \ref{pe-ax}), where no additive penetration was observed. Figure \ref{ternarios7}a shows the $C$--$C_g$ RDFs for both the binary PE-additive A and ternary PE-additiveA-excess water systems. The difference between the two pair correlation functions is clear: the interactions between these atom types are attractive when the system contains water and repulsive otherwise. This reflects a higher affinity of the hydrocarbonated chain of A with PE against water when the latter is in excess, and a preferable interaction of the galactose moiety with water at the interface. 

For additive B, the orientation of the molecules changed from the pocket-like structure of galactitol groups to one similar to that of additive A, with their hydroxyl groups interacting with the water molecules.  This appears to have resulted in the B molecules being positioned mainly outside the polymer rather than being partially embedded as seen in the binary PE-B system. The $C_g$--$O_w$ RDF is shown in Figure \ref{ternarios7}b, which illustrates that this interaction is attractive only for additive B. This is consistent with the presence of chains of additive B at the surface, interacting both with water and PE. In contrast, the hydrocarbon chains of A and EA remain immersed in the polymer, far from the water molecules, thus leading to repulsive interactions.

It is worth noting that water molecules shown in Figure \ref{pe-axinwater}a do not cover the entire PE surface as they did in the PE-water binary system, but adopt a drop-like structure. This structure is in contact with the sugar moieties of additive A. Considering that the area of the analyzed surface is reduced (around 4000 \AA$^2$), this result suggests that the presence of this additive can modify PE surface affinity with water, making it more hydrophobic. Although this was more noticeable for additive A, a drop-like structure was also observed in some of the additive B trajectories. Thus, the ability of this additive to modify PE surface affinity with water cannot be ruled out. 

A different behavior was observed for EA, whose molecules remained immersed in the polymer with the same orientation seen for the binary system in Figure \ref{pe-ax}, and the polymer acquired an ordered structure. This is typical of migratory amphiphilic additives\cite{butuc2017antistatic}, as they are expected to be akin to the polymer but also able to attract moisture from the environment in the polymeric surface through their polar heads. Moreover, Figure \ref{pe-axinwater}c shows that EA molecules cover the entire PE surface homogeneously, and this was consistently observed in the five trajectories. The ability of this additive to make the water molecules disperse in this way over the PE surface agrees with the proposed mechanism of action of antistatic agents in polyolefins, which form a conductive thin layer that dissipates the electrical charges\cite{grossman1993antistatic, butuc2017antistatic}.

Regarding hydrogen bonds, Figure \ref{ternarios7}c shows the same trend as in the binary additive-water cases: this interaction is preferably exhibited between $O_{w}$ and $H_{o}$ for additives A and B, while with EA, hydrogen bonds are established between $O_{h}$ and $H_{w}$ atoms. The $N_{T}-H_{w}$ hydrogen bond is still less favored than that between $O_{h}-H_{w}$, as shown in Figure \ref{ternarios7}d. The $O_{w}$--$H_{o}$ interaction is also possible (attractive) for EA, but it is established at around 3.5 \AA, a larger distance than for the A and B cases (around 1.82 \AA). This was also observed in the binary additive-water systems with excess water analyzed above (Figure \ref{hbonds-axenagua}), and could be interpreted as a way of optimizing the interactions of EA molecules with water through their polar heads while their hydrocarbon tails are immersed in PE. All these results suggest that the additives adapt their structural arrangements to optimize interactions with both the polymer and water simultaneously. The configurations shown in Figure \ref{pe-axinwater} meet this purpose. 

In addition, in replicas where extended simulation times were required to reach equilibrium, the PE matrix exhibited a highly ordered structure, with chains clearly aligning along a preferential axis. This behavior is particularly evident for additives B and EA, as illustrated in the snapshots in Figure S8. Consequently, with sufficient dynamics time, the polymer may attain a highly ordered, crystalline-like configuration.

\begin{figure}[H]
 	\includegraphics[scale=0.82]{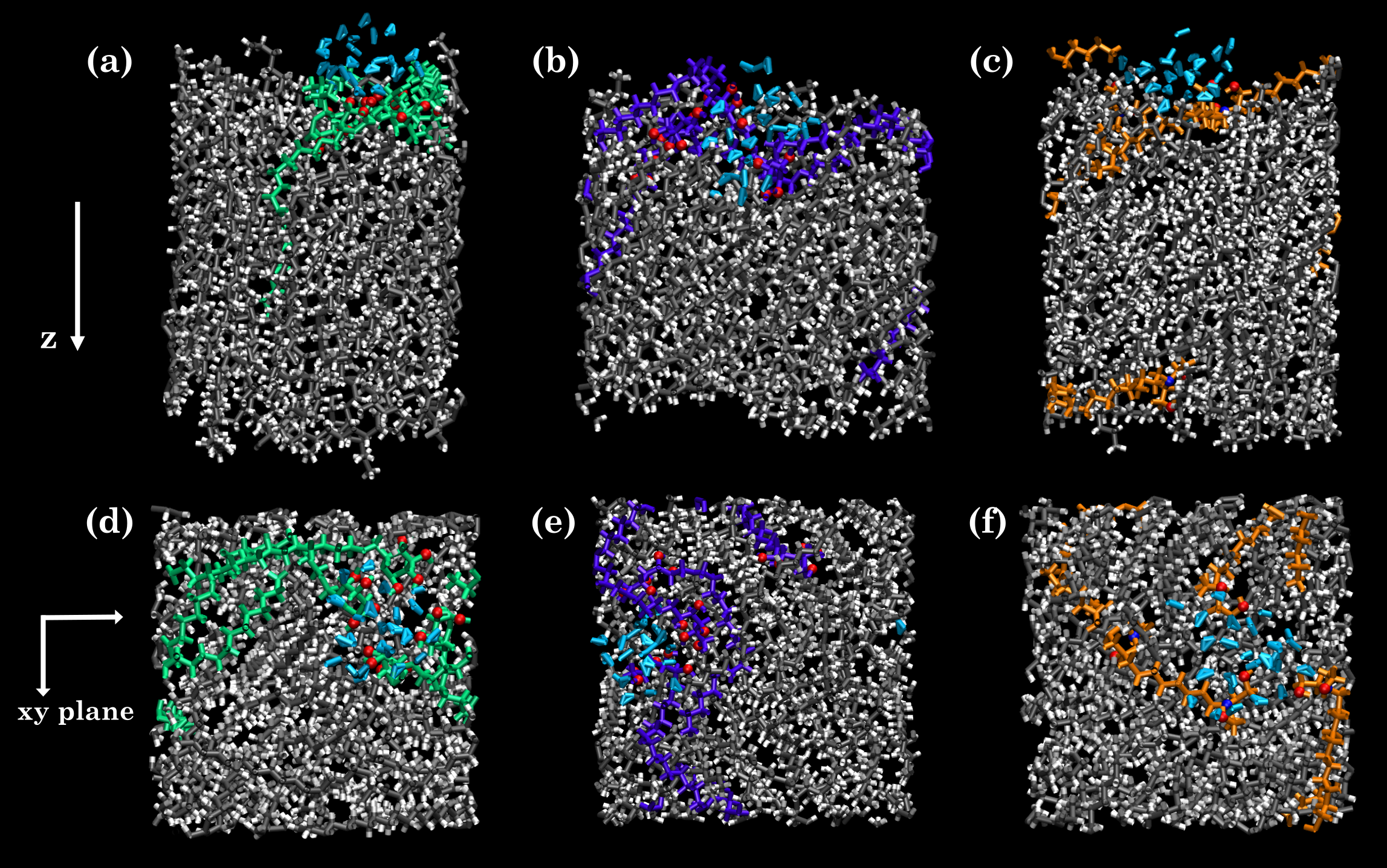}
	\centering
    	 \caption[] {Snapshots of the ternary PE-additive-25-water systems with \textbf{(a)} additive A; \textbf{(b)} B and \textbf{(c)} EA, all of them oriented as in Figure \ref{pe-axinwater}. From \textbf{d} to \textbf{f}, the xy planes of each system are shown.}  
	\label{newtern}
 \end{figure}

To see the effect of water concentration on the structure of PE-additive-excess water systems, five independent runs were performed by reducing the number of water molecules from 220 to only 25. Figure \ref{newtern} is the resulting analog to  Figure \ref{pe-axinwater} for these systems.
The first difference between these ternary systems concerns the structure of water molecules at the surface. While with 220 waters, the surface looked "flooded" and the coverage seemed to be homogeneous, water adopts a drop-like disposition in the 25 water molecules case instead. This water drop is located above the polar moieties of the additives. From this observation, it can be inferred that in the system with 220 water molecules, the water-polar moieties interactions are established until all available hydrogen-bonding sites reach saturation, which leaves several water molecules interacting with other water molecules and with the hydrocarbonated groups of PE and additives. In contrast, when there are only 25 water molecules, the water-polar moieties interactions are enough to keep a drop-like disposition.  

Another difference observed when comparing Figure \ref{newtern} with the snapshots shown in Figure \ref{pe-axinwater} is that the additives look less immersed in the PE matrix when fewer water molecules are in the system.

\begin{figure}[H]
 	\includegraphics[scale=0.82]{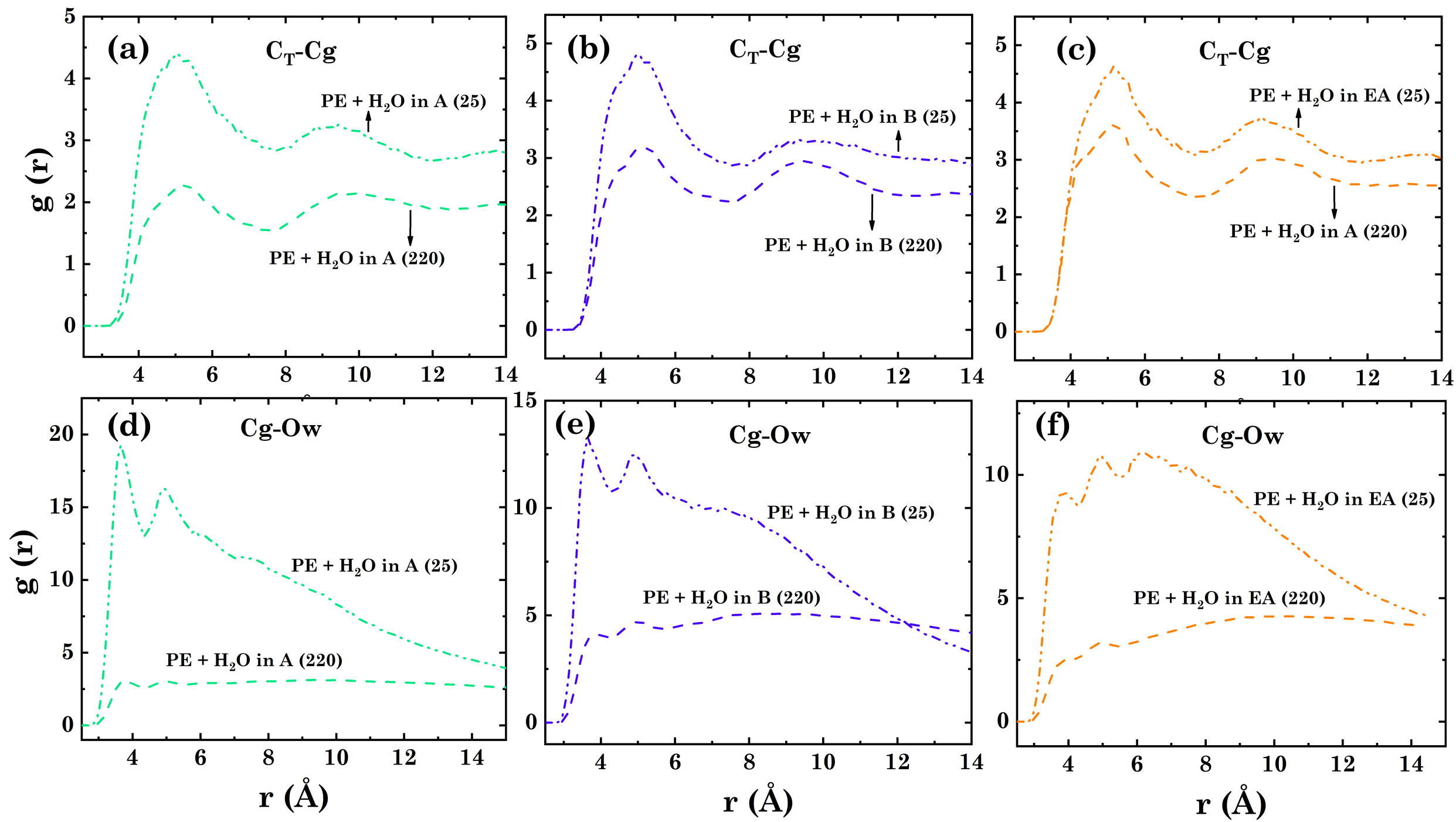}
	\centering
    	 \caption[] {Averaged $C_T$--$C_g$ RDFs of the ternary PE-additive-25-waters systems (dashed-dotted lines labeled as PE + $H_2O$ in X (25)) and the PE-additive-220-waters systems (dashed lines labeled as PE + $H_2O$ in X (220)), where X is \textbf{(a)} additive A; \textbf{(b)} B and \textbf{(c)} EA, and averaged $C_g$--$O_w$ RDFs (from \textbf{d} to \textbf{f}) for systems with additives A, B, and EA, respectively.}  
	\label{newfig}
 \end{figure}

This is clearly reflected in the $C_T$--$C_g$ and $C_g$--$O_w$ RDFs shown in Figure \ref{newfig}. With 220 molecules (dashed lines), the $C_T$--$C_g$ interactions (panels a, b, and c) are slightly attractive for the three additives, and their intensities increase from additive A, followed by B, and then by EA. With fewer water molecules (dashed-dotted lines), these interactions become notoriously more attractive in all cases, indicating that the additives are closer to the PE surface. 
The $C_g$--$O_w$ interactions (Figure \ref{newfig}d-f) support this interpretation for additives A and B, as they are repulsive when 220 water molecules are present (as discussed previously), but strongly attractive with fewer water molecules. In the case of EA, the $C_g$--$O_w$ interactions are less defined compared with A and B, which might be due to the additional hydrogen bond that is possible with EA ($N_{T}-H_{w}$), as it provides more interaction sites with water molecules. This reinforces the prediction of higher water-affinity of the ethoxyamine group relative to the sugar-based moieties, as previously discussed.

$C$--$C_g$ interactions, however, seem to remain invariant when the amount of water molecules changes from 220 to 25 (see Figure S9), thus suggesting that in the systems with 25 water molecules, the hydrocarbonated chains of the additives are still immersed in the PE, but closer to the surface than in the case with 220 water molecules, where a higher repulsion to water was observed.       

Overall, these observations indicate that the presence of water modulates the affinity of the additives for PE and induces the systems to acquire different arrangements to optimize polar and non-polar interactions. When the number of water molecules is higher, the non-polar interactions (established between the polymer and the hydrocarbonated chains of the additives), are compensated with the hydrogen bonds between water molecules. When less water is present, the additives tend to establish a balance between interactions with water and PE, which reflects their amphiphilic nature. 
Moreover, the findings suggest a gradient in amphiphilic behavior governed by the affinity balance of the molecular components. Additive B appears as the less amphiphilic additive, as it only slightly immerses in the PE matrix in the binary system, and is even less immersed in the presence of water. In addition, it has the less water-soluble sugar moiety. Conversely, EA demonstrates the most pronounced amphiphilic nature, while additive A sits between these two cases. Achieving an optimal balance between these features is critical when evaluating additives A and B as potential antistatic agents for PE. For EA, these results are consistent with its established use in commercial antistatic formulations for polyolefins\cite{nouryonArmostat1800, googleUS4093676AAntistatic}, and its well-known migratory nature\cite{stipek2012additives}. Its behavior within the polymer aligns with expectations: the hydrocarbon chains are embedded within the PE matrix and the polar groups oriented towards the PE-air interface\cite{butuc2017antistatic}, enabling simultaneous interaction with water and PE\cite{walp2000antistatic}. 

Considering also that additives A and B have a higher molecular weight (432.6 g/mol and 434.6 g/mol, respectively) than EA (357.6 g/mol), the trend in headgroup polarity and the molecular weights indicate that additives A and B might have a reduced migratory capacity; however, this hypothesis would require experimental verification by incorporating the additives into the bulk PE bulk and monitoring their potential migratory behavior by measuring surface resistivity and coefficients of friction of the surfaces within time.  

\begin{figure}[H]
 	\includegraphics[scale=0.8]{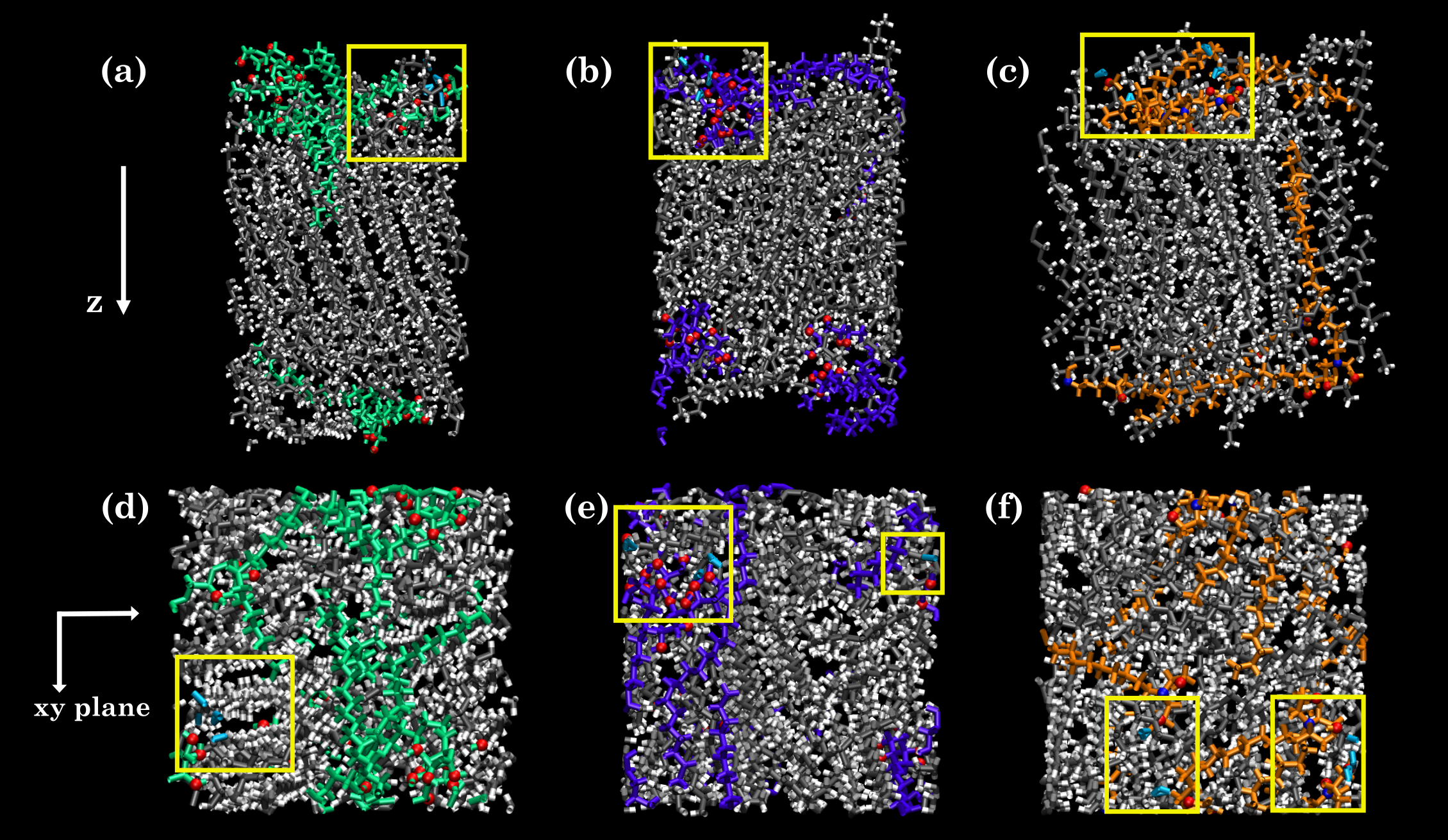}
	\centering
    	 \caption[] {Snapshots of the PE-water-excess additive systems with \textbf{(a)} additive A; \textbf{(b)} B and \textbf{(c)} EA, all of them oriented as in Figure \ref{pe-axinwater}. From \textbf{d} to \textbf{f}, the xy planes of each system are shown. Yellow boxes point to the location of the few water molecules present (coloured in light blue).}  
	\label{pe-waterinax}
 \end{figure}

Finally, \textbf{PE-water-excess additive} ternary systems were investigated. Figure \ref{pe-waterinax} is analogous to Figure \ref{pe-axinwater}, with the simulation boxes oriented in the same way. 

 All the systems have similar structures to those of their analogue binary ones (without water, see Figure \ref{pe-ax}). The only differences are that in the ternary systems the hydrocarbon chains of additive A go slightly deeper in the PE matrix in the ternary system, and that EA molecules are more immersed in the polymer and aligned with the hydrocabonated chains. The similarity was expected given that the systems are nearly identical, differing only by a few water molecules in one of them, highlighted by the yellow squares in Figure \ref{pe-waterinax}. This reinforces the idea that the PE-additive binary systems can be considered equivalent to the real scenario with low water content from the environmental moisture. These water molecules spend most of the time close to the sugar-based and ethoxyamine moieties, as expected, and no significant difference is observed in these interactions between the three additives. 

 The slight differences observed for A and EA reinforce the idea pointed out before regarding the amphiphilic balance of each additive: in the presence of a small amount of water, EA exhibits the best equilibrium between interacting with water and the polymer, followed by additive A (which remains slightly immersed in the PE), and then by B, which seems to behave in a similar way than in the complete absence of water. This behavior is consistent with the water-affinity trend previously described for the sugar and ethoxyamine groups, which decreases from EA to additives A and B. Consequently, the more water-affine the additive (EA), the more its ability to rearrange its interfacial configuration in the presence of water to balance the interactions with both the aqueous phase and the PE matrix. In contrast, for the less water-affine additive (B), the presence of water has a negligible effect, as the additive preferentially maintains its interactions with the PE. 
 
\section{Conclusions}

In this contribution, two organic molecules (labeled as A and B, see Figure \ref{fig1}) were explored as prospective additives for polyethylene industrial formulations via classical atomistic molecular dynamics simulations. To compare the behavior of these additives with a reference example and validate the model, a commercial additive (EA) was also investigated.

Simulations of the pure components confirmed that the employed force fields were appropriate for capturing structural properties (densities and interatomic distances) with reasonable accuracy. Binary additive-water simulations show that all the additives tested are predominantly hydrophobic molecules, as they tend to form aggregates in water. A and B form bilayer-like aggregates, while in EA the molecules align parallel. When the additives are the solvent, water molecules do not aggregate within the simulation times explored. This suggests that the water/additives phase diagram is strongly dependent on relative concentration. The most favorable interactions between water and the additives are hydrogen bonds established with the polar regions of the additives, with water acting preferentially as a hydrogen bond acceptor when combined with A or B and as a donor when combined with EA. 

The polyethylene slab models exhibit a crystalline-like center and amorphous-like surfaces. Polymer terminations tend to accumulate at the surface of the slab, in agreement with previous computational works. 

Binary polyethylene-water simulations confirm the low water-polymer affinity, with water molecules covering the polymeric surface but never entering its matrix. No preferential orientation was found for the water molecules. Concerning polymer-additive binary systems, a trend of increasing polymer affinity was found when going from B to A and EA. Indeed, the ethoxyamine polar heads of EA molecules stay available at the PE surface, while the polar heads of A and B cluster together, with polar heads in additive B going deeper into the non-polar polymer matrix. The structure exhibited by EA is in agreement with the migratory behavior previously identified for this commercial additive, while the structure of additive B suggests that it could be considered as a potential slip agent for polyethylene. 

The structure of ternary systems containing PE, water, and additives strongly depends on the relative concentrations of the components. For systems containing low additive concentrations and 220 water molecules (enough to form a water layer over the surface), water homogeneously covers the polyethylene surface. The structure of the additives remains quite unaltered in this scenario with respect to the binary analogs, except for additive A, which gets further into the polyethylene matrix. Since water covers the polymer surface homogeneously, interactions with the polar heads and non-polar tails of the additives and with polyethylene are all observed. When the water concentration is reduced (25 water molecules), water adopts a drop-like structure instead, and only hydrogen bonds with the polar heads of the additives are observed. Additives A and EA penetrate less deeply into the polymer matrix in this case, while additive B remains unaffected. Finally, PE-water-excess additive systems look very similar to the PE-additive binary systems. These systems are the best model for understanding the structure of the additives in industrial PE films exposed to air humidity at ambient temperature.  

This work provides insights into the interfacial structure of a series of additives in polyethylene formulations and water. Results suggest that the interplay of water/additive interactions is strongly correlated to the relative ratio, while those with the polymer are less impacted, except for the case of additive A. In all ternary systems, the additives adapt their disposition to maximize both their interactions with water and polyethylene, due to their amphiphilic character. Water adapts its structure to minimize contact with polyethylene while maximizing interactions with other water molecules and the polar heads of the additives. Our results provide consistent molecular-scale evidence of the proposed working mechanism of EA as an antistatic agent\cite{walp2000antistatic, butuc2017antistatic}, and suggest that it might be interesting to further explore additive B for its potential slip capacity. We hope that these studies will motivate further experimental efforts in this direction.


\begin{acknowledgments}

MMC thanks for a CONICET-Ampacet South America S.R.L PhD fellowship. RMN is a member of the Carrera de Investigador Científico of CONICET (Argentina). RS thanks the European Research Council for an ERC StG (MAGNIFY project, number 101042514). RMN and RS are grateful to the Comité ECOS Sud (France), for the ECOS Sud 2020 A20N04 project, which has enabled this collaboration.  

\end{acknowledgments}

\section*{Data availability}

The data supporting this article have been included as part of the Supporting Information (SI). SI includes force field parameters and additional results that support those presented in the article.  

\section*{Conflicts of Interests}

The authors have no conflicts of interest to declare.

\section{Supporting Information}

\subsection{Force Field}

\begin{figure}[H]
 	\includegraphics[scale=0.85]{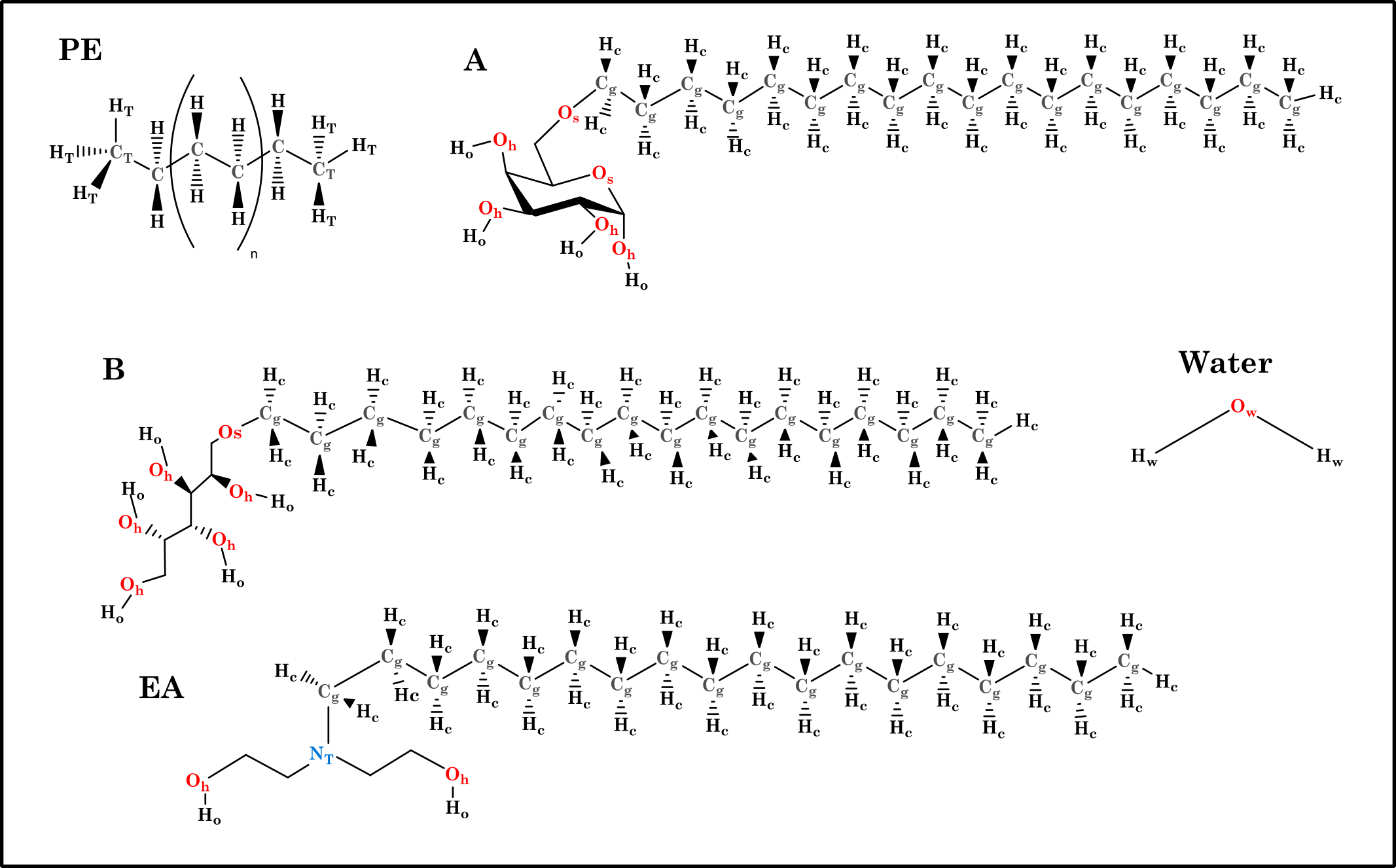}
	\centering
    	 \caption[g de r agua en ax] {Chemical structures of the simulated molecules: PE, additives A, B, and EA, and water. Relevant atom types are labelled: $C_{g}$ and $H_{c}$ are the atoms of the hydrocarbon chains, $O_{h}$  and $H_{o}$ the oxygen and hydrogen atoms in the hydroxyl groups of the sugar moieties, $N_{T}$ the nitrogen atom of EA, and $O_{w}$ and $H_{w}$ the water atom types. O, C, N and H atoms are represented in red, grey, blue, and black, respectively. In the PE structure, subfix \textit{n} indicates the number of monomers contained in the chain, $15\leq n \leq 48$.}  
	\label{atomtypes}
 \end{figure}

\newpage

 Tables \ref{ffaditivos}, \ref{ffpolymer}, and \ref{ffwater} show the force field parameters for all species. In all cases, dihedrals were described in the harmonic style. Cross interactions between and within phases were computed by applying the Lorentz-Berthelot mixing rules.

\begin{center}
\begin{longtable}{cccc}
\caption{Force fields parameters for the additives extracted from GLYCAM06.} \label{ffaditivos} \\
\toprule
\textbf{Non-bonded terms} & \textbf{$\epsilon_{ii}$ (kcal/mol)} & \textbf{$\sigma_{ii}$ (\AA)} & \textbf{$q_{i}$ (e)} \\
\midrule
\endfirsthead

\endhead

\bottomrule
\endlastfoot

$C_{g}$ & 0.1094 & 3.3996 & -0.23006 \\
3C & 0.1094 & 3.3996 & 0.301333 \\
$H_{o}$ & 0.03 & 0.3563 & 0.390500 \\
$H_{1}$ & 0.0157 & 2.4713 & 0.000000 \\
$H_{2}$ & 0.0157 & 2.2932 & 0.000000 \\
$H_{c}$ & 0.0157 & 2.6495 & 0.118000 \\
$O_{h}$ & 0.2104 & 3.0064 & -0.644000 \\
$O_{s}$ & 0.1700 & 3.0000 & -0.606667 \\
$N_{T}$ & 0.1700 & 3.2500 & -0.938000 \\
\midrule
\textbf{Bonds} & \textbf{$K_{r}$ (kcal/(mol\AA$^2$))} & \textbf{$r_{eq}$ (\AA)} & \\
$3C - O_{h}$ & 320 & 1.43 & \\
$3C - O_{s}$ & 285 & 1.46 & \\
$O_{h} - H_{o}$ & 553 & 0.96 & \\
$C_{g} - C_{g}$ & 310 & 1.52 & \\
$3C - C_{g}$ & 310 & 1.52 & \\
$3C - H_{2}$ & 340 & 1.09 & \\
$3C - H_{1}$ & 340 & 1.09 & \\
$C_{g} - H_{c}$ & 340 & 1.09 & \\
$3C - N_{T}$ & 352 & 1.47 & \\
\midrule
\textbf{Angles} & \textbf{$K_{\theta}$ (kcal/(mol rad))} & \textbf{$\theta_{eq}$ (\textdegree)} & \\
$3C-O_{h}-H_{o}$ & 55 & 109.5 & \\
$O_{s}-3C-H_{1}$ & 60 & 110 & \\
$O_{h}-3C-H_{1}$ & 60 & 110 & \\
$O_{s}-3C-H_{2}$ & 60 & 110 & \\
$O_{h}-3C-H_{2}$ & 60 & 110 & \\
$C_{g}-C_{g}-C_{g}$ & 45 & 113.5 & \\
$3C-O_{s}-3C$ & 50 & 111.6 & \\
$3C-3C-O_{s}$ & 70 & 108.5 & \\
$C_{g}-3C-O_{s}$ & 70 & 108.5 & \\
$3C-3C-O_{h}$ & 70 & 107.5 & \\
$H_{c}-C_{g}-H_{c}$ & 40 & 109.5 & \\
$H_{1}-3C-H_{1}$ & 45 & 109.5 & \\
$C_{g}-C_{g}-H_{c}$ & 45 & 112.6 & \\
$C_{g}-3C-H_{1}$ & 45 & 111 & \\
$3C-3C-H_{1}$ & 45 & 111 & \\
$3C-3C-H_{2}$ & 45 & 111 & \\
$O_{s}-3C-O_{h}$ & 100 & 112 & \\
$H_{1}-3C-N_{T}$ & 64 & 109.6 & \\
$3C-N_{T}-3C$ & 54 & 108.1 & \\
$C_{g}-3C-N_{T}$ & 67 & 111.6 & \\
$3C-3C-N_{T}$ & 67 & 111.6 & \\
\midrule
\textbf{Dihedrals} & \textbf{$V_{n}/2$ (kcal/mol)} & \textbf{d} & \textbf{n} \\
$H_{o} - O_{h} - 3C - H_{1}$ & 0.18 & 1 & 3 \\
$H_{o} - O_{h} - 3C - H_{2}$ & 0.18 & 1 & 3 \\
$O_{h} - 3C - 3C - H_{1}$ & 0.05 & 1 & 3 \\
$O_{s} - 3C - 3C - H_{1}$ & 0.05 & 1 & 3 \\
$O_{s} - 3C - C_{g} - H_{c}$ & 0.05 & 1 & 3 \\
$O_{h} - 3C - 3C - H_{2}$ & 0.05 & 1 & 3 \\
$H_{c} - C_{g} - C_{g} - H_{c}$ & 0.13 & 1 & 3 \\
$H_{1} - 3C- 3C - H_{1}$ & 0.17 & 1 & 3 \\
$H_{c} - C_{g} - 3C - H_{1}$ & 0.17 & 1 & 3 \\
$H_{1} - 3C- 3C - H_{2}$ & 0.17 & 1 & 3 \\
$H_{c} - C_{g} - C_{g} - C_{g}$ & 0.1 & 1 & 3 \\
$H_{1} - 3C - C_{g} - H_{c}$ & 0.15 & 1 & 3 \\
$H_{2} - 3C- 3C - 3C$ & 0.15 & 1 & 3 \\
$C_{g} - C_{g} - C_{g} - C_{g}$ & 0.45 & 1 & 1 \\
$O_{h} - 3C- 3C - 3C$ & 0.1 & 1 & 3 \\
$O_{s} - 3C- 3C - 3C$ & -0.27 & 1 & 1 \\
$O_{s} - 3C- C_{g} - C_{g}$ & -0.27 & 1 & 1 \\
$3C - 3C- O_{h} - H_{o}$ & 0.18 & 1 & 3 \\
$O_{h} - 3C- 3C - O_{h}$ & -0.1 & 1 & 1 \\
$O_{s} - 3C- 3C - O_{h}$ & -1.1 & 1 & 1 \\
$O_{s} - 3C- 3C - O_{s}$ & 0.40 & 1 & 2 \\
$3C - 3C- O_{s} - C_{g}$ & 0.16 & 1 & 3 \\
$3C - O_{s} - 3C - C_{g}$ & 0.16 & 1 & 3 \\
$3C - O_{s} - 3C - H_{1}$ & 0.27 & 1 & 3 \\
$3C - O_{s} - 3C - H_{2}$ & 0.60 & 1 & 2 \\
$O_{h} - 3C- O_{s} - 3C$ & 0.96 & 1 & 3 \\
$H_{o} - O_{h} - 3C - O_{s}$ & 0.18 & 1 & 3 \\
$H_{1} - 3C- N_{T} - 3C$ & 0.25 & 1 & 3 \\
$3C - N_{T} - 3C - C_{g}$ & 0.1 & -1 & 3 \\
$C_{g} - C_{g} - 3C - N_{T}$ & 0.3 & -1 & 3 \\
$O_{h} - 3C- 3C - N_{T}$ & 0.6 & -1 & 3 \\
$H_{1} - 3C- 3C - N_{T}$ & 0.1 & 1 & 3 \\
$H_{c} - C_{g}- 3C - N_{T}$ & 0.1 & 1 & 3 \\
\end{longtable}
\end{center}

\begin{table}[H]
\centering
\caption{Force field parameters for polyethylene extracted from GAFF. }
\label{ffpolymer}
\begin{tabular}{@{}cccc@{}}
\toprule
\textbf{Non-bonded terms}      & \textbf{$\epsilon_{ii}$ (kcal/mol)} & \textbf{$\sigma_{ii}$ (\AA)} & \textbf{$q_{i}$ (e)} \\ \midrule
C                       & 0.1094                              & 3.39967                      & -0.2043716           \\
$C_{T}$                 & 0.1094                              & 3.39967                      & -0.2043716           \\
H                       & 0.0157                              & 2.64963                      & 0.1000000            \\
$H_{T}$                 & 0.0157                              & 2.64963                      & 0.1000000            \\ \hline
\textbf{Bonds}           & \textbf{$K_{r}$ (kcal/(mol\AA$^2$))}         & \textbf{$r_{eq}$ (\AA)}      & \textbf{}            \\
C - C                   & 303.1                               & 1.535                        &                      \\
$C - C_{T}$             & 303.1                               & 1.535                        & \multicolumn{1}{l}{} \\
H - C                   & 337.3                               & 1.092                        &                      \\
$H_{T} - C_{T}$         & 337.3                               & 1.092                        &                      \\ \hline
\textbf{Angles}          & \textbf{$K_{\theta}$ (kcal/(mol rad))}    & \textbf{$\theta_{eq}$ (\textdegree)} &                      \\
C-C-C                   & 63.21                               & 110.63                       &                      \\
$C_{T}-C-C$             & 63.21                               & 110.63                       & \multicolumn{1}{l}{} \\
H-C-C                   & 46.37                               & 110.05                       & \multicolumn{1}{l}{} \\
$H-C-C_{T}$             & 46.37                               & 110.05                       & \multicolumn{1}{l}{} \\
$H_{T}-C_{T}-C$         & 46.37                               & 110.05                       & \multicolumn{1}{l}{} \\
H-C-H                   & 39.43                               & 108.35                       & \multicolumn{1}{l}{} \\
$H_{T}-C_{T}-H_{T}$     & 39.43                               & 108.35                       & \multicolumn{1}{l}{} \\ \hline
\textbf{Dihedrals}       & \textbf{$V_{n}/2$ (kcal/mol)}                  & \textbf{d}                   & \textbf{n}           \\
C-C-C-C                 & 0.18                                & 1                            & 3                    \\
$C -C -C -C_{T}$        & 0.18                                & 1                            & 3                    \\
$C - C - C_{T} - H_{T}$ & 0.16                                & 1                            & 3                    \\
C-C-C-H                 & 0.16                                & 1                            & 3                    \\
$C_{T} - C - C - H$     & 0.16                                & 1                            & 3                    \\
$H_{T} - C_{T} - C - H$ & 0.15                                & 1                            & 3                    \\
H-C-C-H                 & 0.15                                & 1                            & 3                    \\ \bottomrule
\end{tabular}
\end{table}

\begin{table}[H]
\centering
\caption{Force field parameters for water.}
\label{ffwater}
\begin{tabular}{@{}cccc@{}}
\toprule
\textbf{Non-bonded terms}  & \textbf{$\epsilon_{ii}$ (kcal/mol)} & \textbf{$\sigma_{ii}$ (\AA)} & \textbf{$q_{i}$ (e)} \\ \midrule
$O_{w}$             & 0.152                               & 3.15075                      & -0.834               \\
$H_{w}$             & 0.000                               & 1.79605                      & 0.417                \\ \hline
\textbf{Bond}       & \textbf{$K_{r}$ (kcal/(mol\AA$^2$))}         & \textbf{$r_{eq}$ (\AA)}      & \textbf{}            \\
$O_{w} - H_{w}$     & 553                                 & 0.9572                       &                      \\ \hline
\textbf{Angle}      & \textbf{$K_{\theta}$ (kcal/(mol rad))}    & \textbf{$\theta_{eq}$ (\textdegree)} &                      \\
$H_{w}-O_{w}-H_{w}$ & 100                                 & 104.52                       & \multicolumn{1}{l}{}
\end{tabular}
\end{table}

\newpage

\section{Polyethylene model generation}

The partial atomic charges in the final polymer were of $q_H=$0.1 and $q_C=$-0.2044 (elementary charge units). Initial coordinates of the monomer were obtained from the editor and viewer \textit{Avogadro}\cite{hanwell2012avogadro}.

The polydisperse non-branched polyethylene model obtained contained a total of 24 chains with different lengths, detailed by the number of monomers and total carbon atoms in each of them in Table \ref{tablePE}. 

\begin{table}[H]
\centering
\caption{Description of the chains in the modelled polyethylene by the number of monomers, total carbon atoms, and molecular weight.}
\label{tablePE}
\setlength{\tabcolsep}{1em}
\begin{tabular}{cccc} 

\toprule

\textbf{Number of} & \textbf{Total carbon} & \textbf{$M_w$ per}  & \textbf{Number of} \\[-10pt]

\textbf{monomers} & \textbf{atoms} & \textbf{chain (g/mol)} &  \textbf{chains} \\

\midrule
15                          & 30                                              & 422                          & 2                         \\
16                          & 32                                              & 450                          & 2                         \\
17                          & 34                                              & 478                          & 2                         \\
18                          & 36                                              & 506                          & 1                         \\
19                          & 38                                              & 534                          & 2                         \\
20                          & 40                                              & 562                          & 3                         \\
21                          & 42                                              & 590                          & 1                         \\
22                          & 44                                              & 618                          & 2                         \\
23                          & 46                                              & 646                          & 3                         \\
24                          & 48                                              & 674                          & 1                         \\
27                          & 54                                              & 758                          & 1                         \\
32                          & 64                                              & 898                          & 1                         \\
34                          & 68                                              & 954                          & 1                         \\
38                          & 76                                              & 1066                         & 1                         \\
48                          & 96                                              & 1346                         & 1                        
\end{tabular}
\end{table}

 \begin{figure}[H]
 	\includegraphics[scale=0.8]{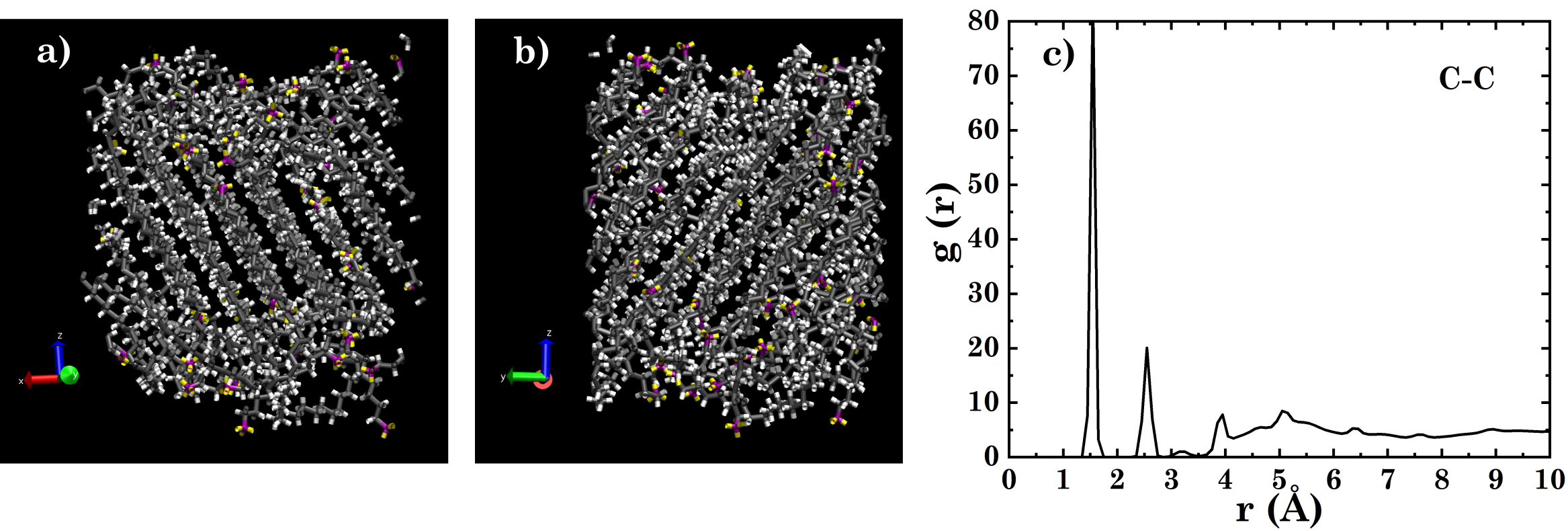}
	\centering
    	 \caption[g de r agua en ax] {\textbf{a)} and \textbf{b)}: Pure polyethylene box seen from two different planes to show the alignment of the chains. Methyl chain ends are highlighted with carbon atoms in purple and hydrogen atoms in yellow. \textbf{c)}: radial pair distribution function for the C-C atom types in the polymer. Results are shown in the same way as those obtained by Narten,\cite{Narten1989} to faciliate seeing the agreement between them.}  
	\label{polymer}
 \end{figure}

  \begin{figure}[H]
 	\includegraphics[scale=0.4]{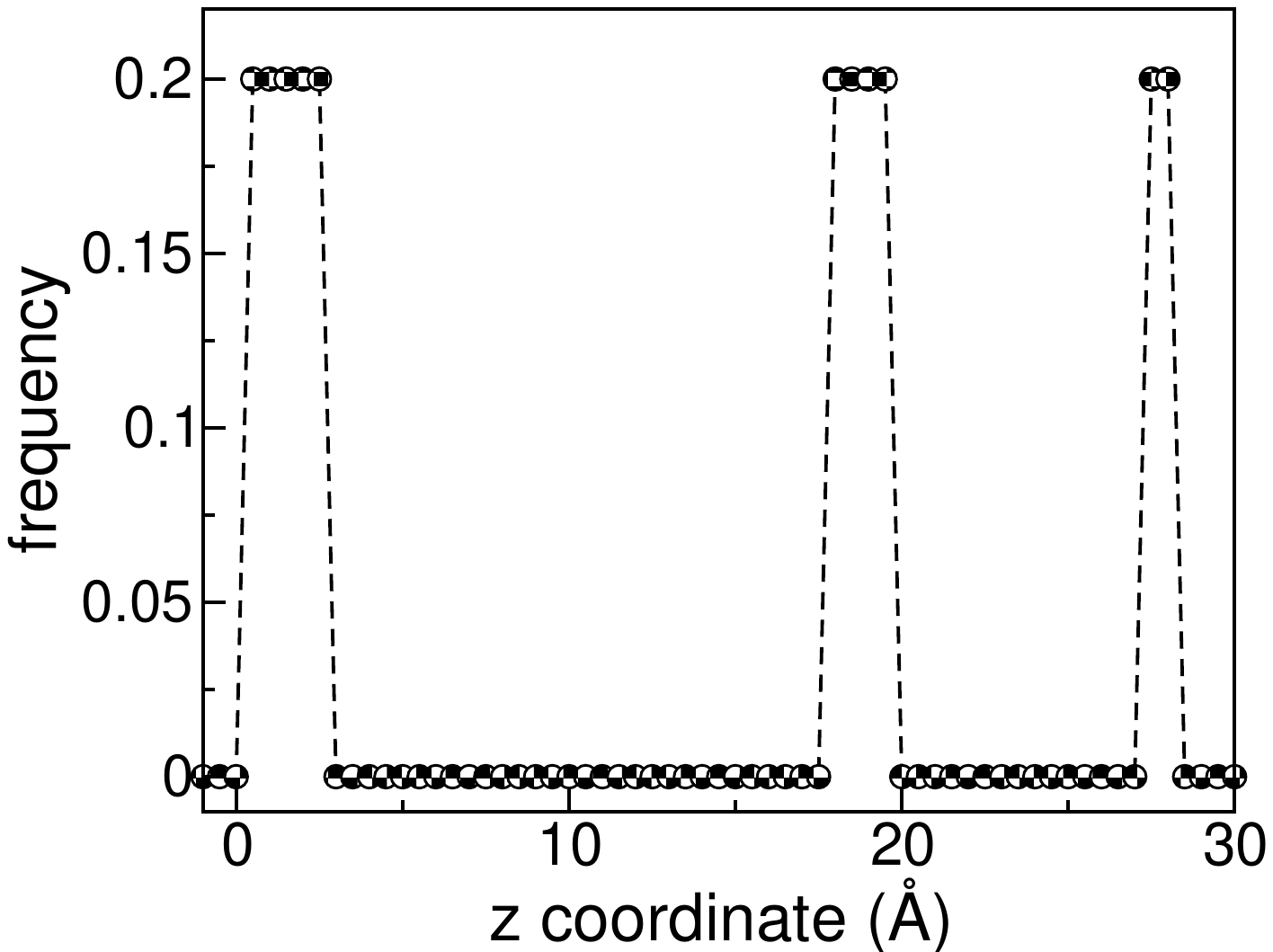}
	\centering
    	 \caption[] {Frequency of appearance of the chain-end carbon atoms of the \textbf{pure polymer slab} systems as a function of z coordinate, which is the direction perpendicular to the polymer slab. The plot presents the average of multiple independent MD runs. The slab surface at lower z values was assigned a z=0 value in all cases, to facilitate comparison by averaging. The peaks around 1 and 27 \AA \ are located close to the interface regions.}  
	\label{zdens}
 \end{figure}

\section{Binary systems}
\subsection{Additives in water}

\begin{figure}[H]
 	\includegraphics[scale=0.8]{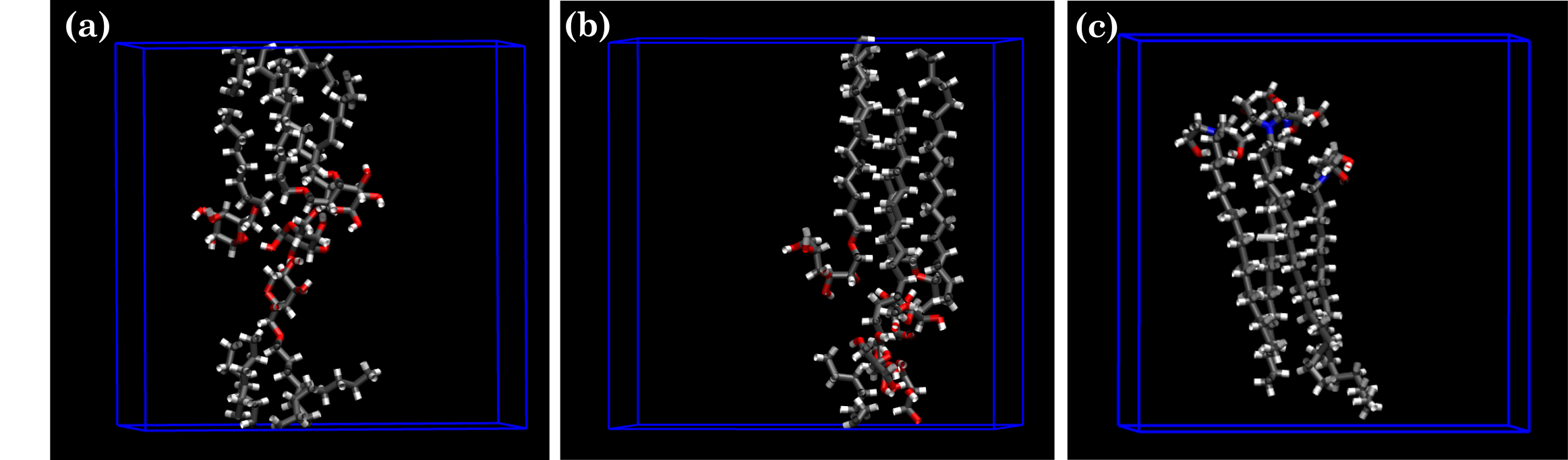}
	\centering
    	 \caption[captures de ax en agua] {Pictures of the \textbf{additives-in-water} binary systems (excess water). Water molecules are not shown for visual clarity. \textbf{a)} Additive A and \textbf{b)} Additive B: in both cases, the additive molecules form aggregates where the polar sugar moieties are in close contact and interact with each other, whereas the hydrocarbon chains are extended and interact among themselves. \textbf{c)} EA: the four additive molecules are together, and their ethoxyamine groups are disposed towards the same place.}  
	\label{axenagua-bin}
 \end{figure}

\begin{figure}[H]
 	\includegraphics[scale=0.8]{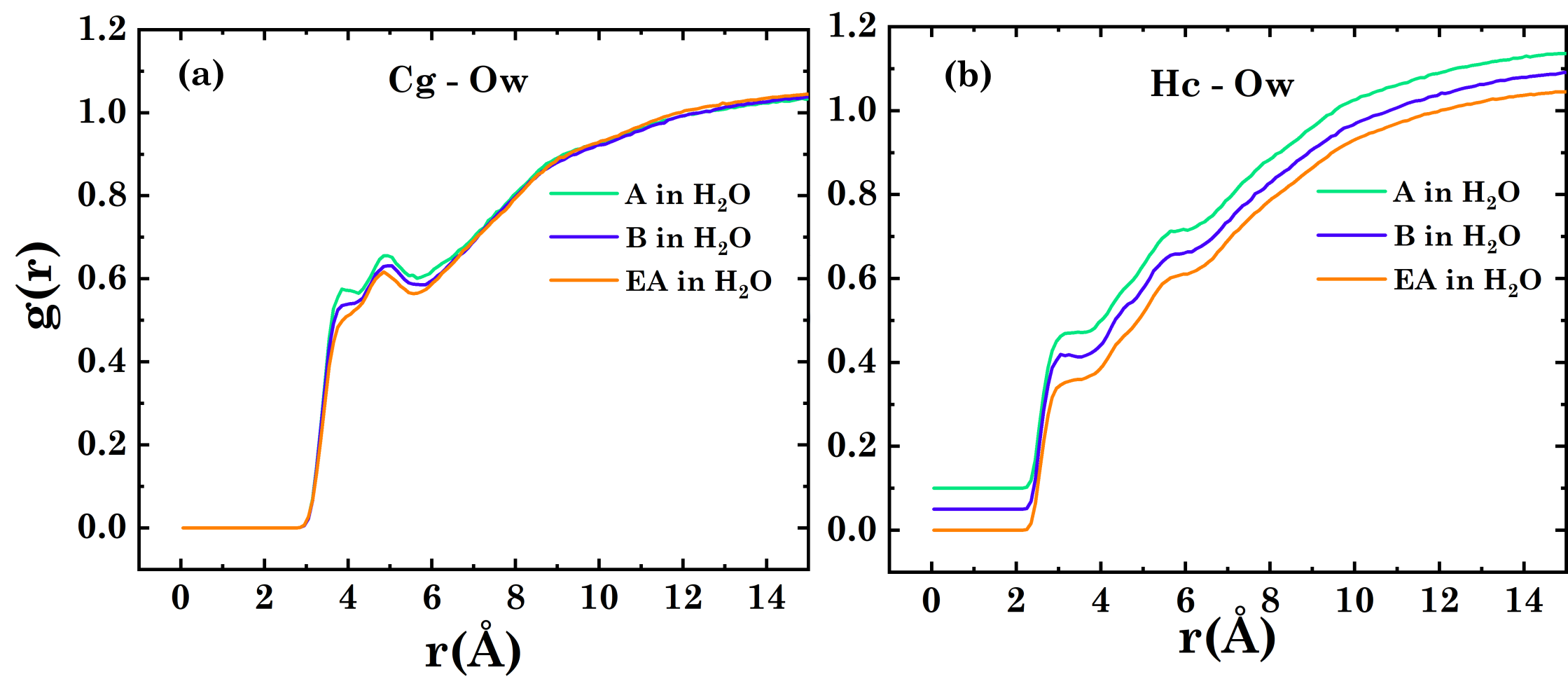}
	\centering
    	 \caption[interac de colas ax en agua] {Averaged RDFs for non-polar interactions between the hydrocarbon chains of the additives and the oxygen atoms of water: \textbf{a)} $C_{g}-O_{w}$ , and \textbf{b)} $H_{c}-O_{w}$; an offset was applied on the y-axis for visual clarity. All the interactions were found to be repulsive.}  
	\label{grcolasaxenagua}
 \end{figure}

 \subsection{Water in additives}

 \begin{figure}[H]  
 \includegraphics[scale=0.8]{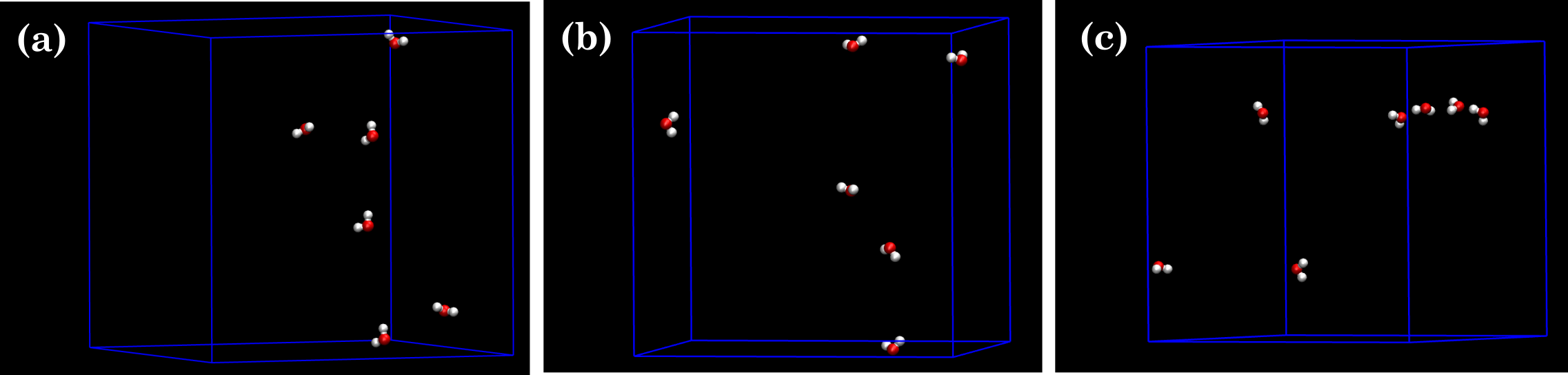}
	\centering
    	 \caption[estrellas de agua] {Snapshots of the \textbf{water-in-additives} systems. Additive molecules are not shown for visual clarity. In the systems with \textbf{a)} additive A, and \textbf{b)} with additive B, water molecules are separated and spread in the box. In the case of \textbf{c)} with EA, some molecules are close to each other, but no clear arrangement can be identified.}  
	\label{estrellasdeagua}
 \end{figure}

 \begin{figure}[H]
    \includegraphics[scale=0.7]{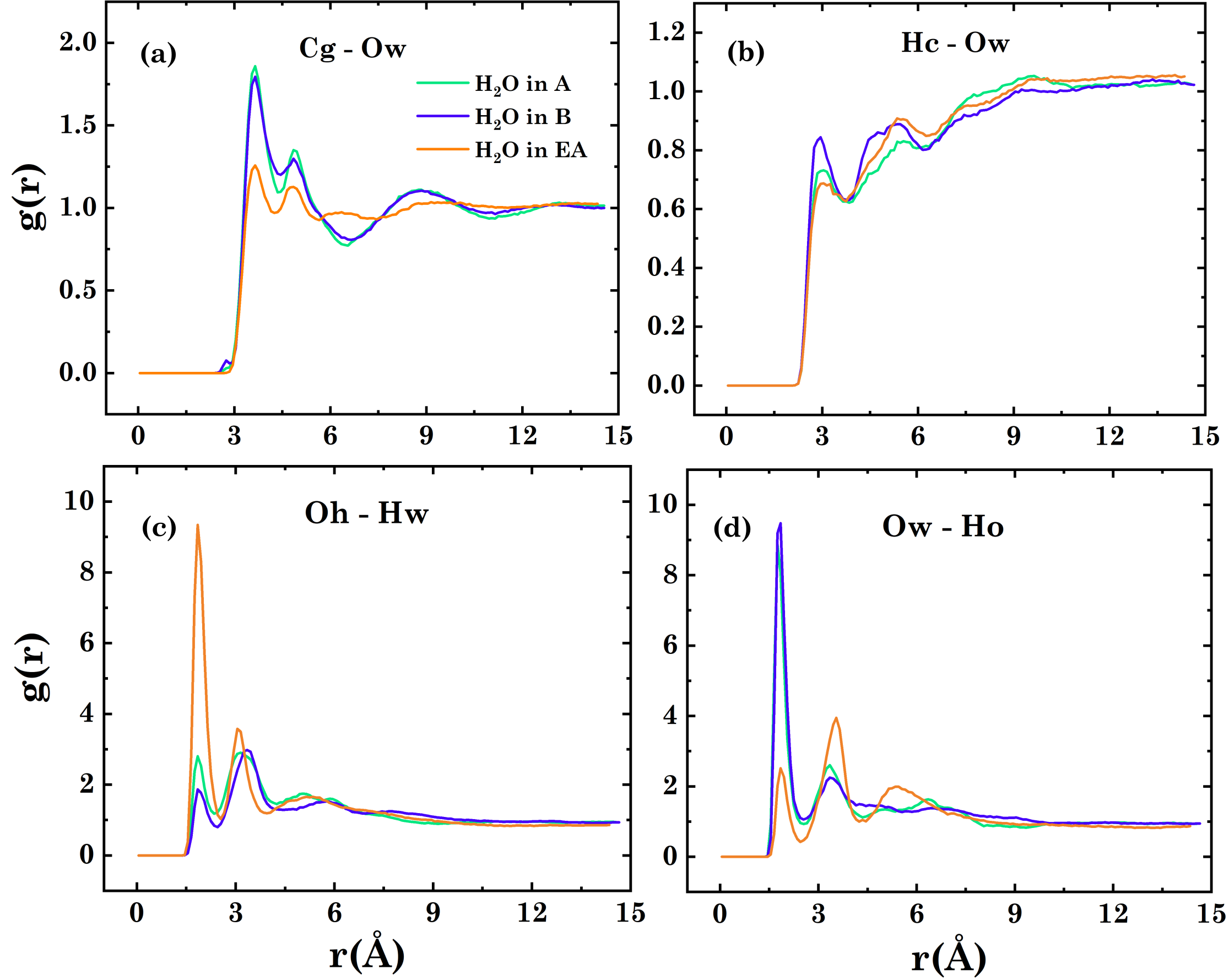}
	\centering
    	 \caption[g de r agua en ax] {RDFs for each \textbf{water-in-additives} system to characterize interactions between water and the hydrocarbon chains of the additives: \textbf{a)} $C_{g}-O_{w}$, \textbf{b)} $H_{c}-O_{w}$, and the hydrogen bonds \textbf{c)} $O_{h}-H_{w}$ and \textbf{d)} $O_{w}-H_{o}$. A clear preference for the $O_{h}-H_{w}$ hydrogen bond can be seen for EA when compared with A and B.}  
	\label{graguaenax}
 \end{figure}

\section{Ternary systems}

 \begin{figure}[H]
 	\includegraphics[scale=0.82]{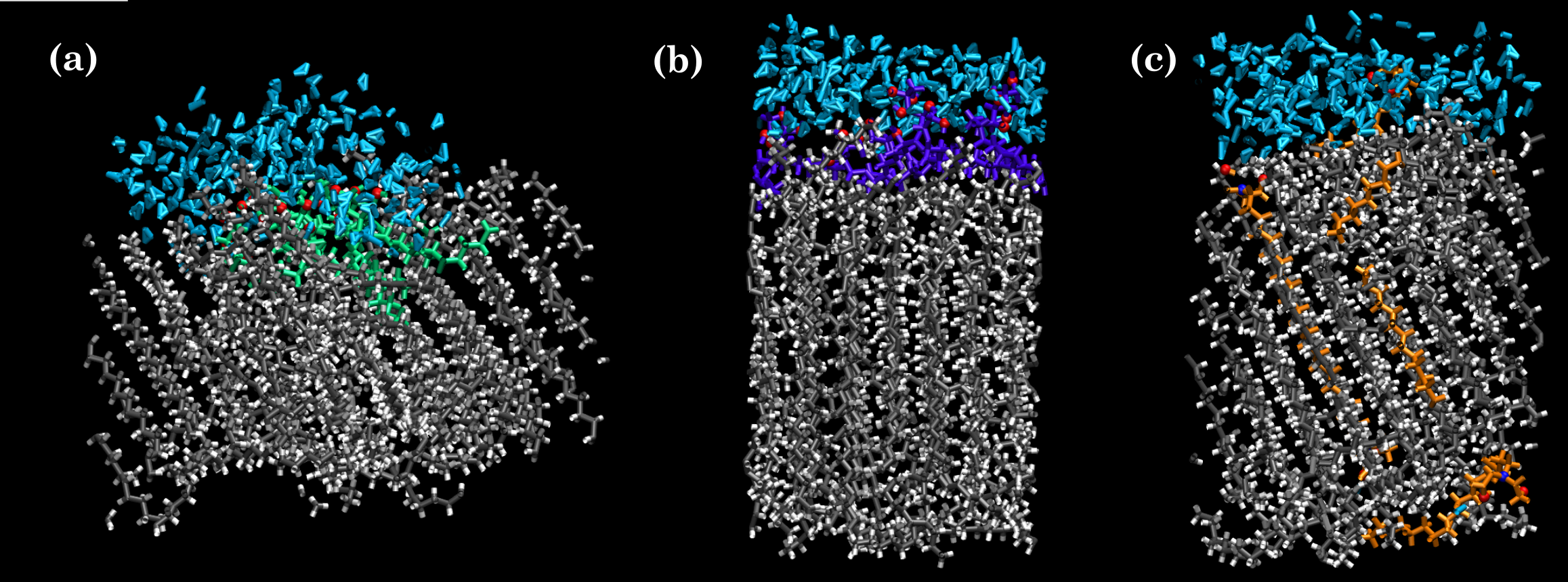}
	\centering
    	 \caption[] {Snapshots of the \textbf{PE-additive-excess water} systems with 220 water molecules for replicas with extended simulation times for \textbf{a)} system with additive A; \textbf{b)} with additive B, and \textbf{c)} with EA.}  
	\label{PE_ordered}
 \end{figure}

 \begin{figure}[H]
 	\includegraphics[scale=0.82]{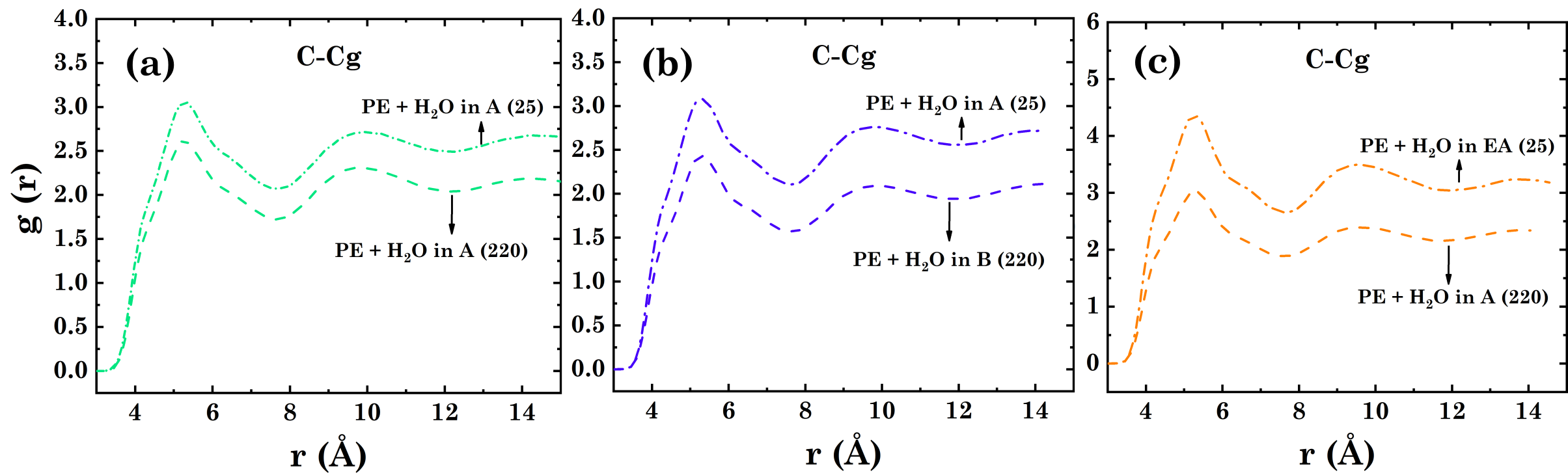}
	\centering
    	 \caption[] {RDFs $C$--$C_{g}$ of the \textbf{PE-additives-excess water} systems with 25 (dash-dot-dot lines) and 220 water molecules (dashed lines) to compare the affinity of the hydrocarbonated chains of additives with PE when the amount of water changes. \textbf{(a)} Systems with additive A, \textbf{(b)} with B, and \textbf{(c)} with EA.}  
	\label{c-cg-ternarios}
 \end{figure}

\newpage

\bibliography{mar}

\end{document}